\documentclass[pre,floatfix,twocolumn]{revtex4}
\usepackage{amsmath}
\usepackage{amsfonts}
\usepackage{mathrsfs}
\usepackage{epsfig}
\usepackage{epstopdf}
\usepackage{hyperref}
\usepackage{graphicx}
\usepackage{dcolumn}
\usepackage{amsmath}
\usepackage{placeins}
\usepackage{xcolor}
\usepackage[utf8]{inputenc} 
\usepackage{longtable}
\usepackage{url}

\def\be{\begin{equation}}
\def\ee{\end{equation}}

\begin{document}
\title{Developmental trajectory of {\em Caenorhabditis elegans} nervous system governs its structural organization}
\author{Anand Pathak$^{1,2}$, Nivedita Chatterjee$^3$ and Sitabhra Sinha$^{1,2}$}
\affiliation{$^1$The Institute of Mathematical Sciences, CIT Campus,
Taramani, Chennai 600113, India.\\
$^2$Homi Bhabha National Institute, Anushaktinagar, Mumbai 400094,
India.\\
$^3$Vision Research Foundation, Sankara Nethralaya, 41 College Road,
Chennai 600006, India.}
\date{\today}
\begin{abstract}
A central problem of neuroscience involves uncovering the principles governing the 
organization of nervous systems which ensure robustness in brain development. 
The nematode {\em Caenorhabditis elegans}
provides us with a model organism for studying this question. 
In this paper, we focus on the invariant connection structure and spatial 
arrangement of the neurons comprising the somatic neuronal network of this organism to 
understand the key developmental constraints underlying 
its design.
We observe that neurons with certain shared characteristics - such
as, neural process lengths, birth time cohort, lineage and bilateral
symmetry - exhibit
a preference for connecting to each other. 
Recognizing the existence of such homophily helps in connecting
the physical location and morphology of neurons with the 
topological organization of the network. Further, 
the functional identities of neurons appear to dictate the
temporal hierarchy of their appearance during the course of development.
Providing crucial insights into principles that may be common across many organisms, 
our study shows how the trajectory in the developmental landscape constrains the eventual spatial 
and network topological organization of a nervous system.
\end{abstract}


\maketitle

\newpage
\section{Introduction}
The presence of an efficient machinery 
for responding immediately to changes in the environment with
appropriate actions is essential for the survival of any
organism. In almost all multicellular animals, this role is played by
the nervous system comprising networks of neurons, specialized cells that 
rapidly exchange signals with a high degree
of accuracy. It allows information about the environment obtained 
via sensory receptors to be processed and translated into output signals
conveyed to effectors such as muscle cells.
In even the simplest of such organisms, the structural description of 
the interconnections between neurons provided by the connectome presents 
an extremely complicated picture~\cite{Lichtman2014}. 
How the complex organization
of the nervous system is generated in the course of development
of an organism, occasionally referred to as the 
``brain wiring problem"~\cite{Hassan2015}, 
is one of the most challenging questions in 
biology~\cite{Mitchell2007,Adolphs2015}.
Only over the past few decades is the intricate interplay of different 
developmental
phenomena, including cellular differentiation, migration, axon guidance 
and synapse formation, responsible for the formation
of the network, being gradually 
revealed~\cite{Araujo2003,Chen2009,ColonRamos2009,Kolodkin2011,Kaiser2017}.

The free-living nematode {\em Caenorhabditis elegans}, the only organism 
whose entire connectome has been reconstructed so far~\cite{White1986,varshney2011}, is the natural
choice for a system in which to look for principles 
governing the development of complexity in the nervous 
system~\cite{Brenner2009}. 
The nervous system of the mature hermaphrodite individuals of the species
comprises $302$ neurons, which is about a third of the total
complement of $959$ somatic cells in the animal. Their lineage, positions in
the body of the worm and connections to each other appear to be almost 
invariant across individuals~\cite{White1986,celegans2}. 
The small number of cells constituting the worm has made it a
relatively tractable system for understanding the genetic basis
of metazoan development and behavior. This, however, belies the 
sophistication of the
organism which exhibits almost all the important specialized tissue
types that occur in larger, more complex animals, 
prompting it to be dubbed as a ``microchip animal''~\cite{Haag2018}.
The availability of its complete genome 
sequence~\cite{Hillier2005} along with detailed information about the
cell lineage~\cite{sulston77,sulston83} means that, in principle,
the developmental program can be understood as a consequence of 
genetically-encoded instructions and
self-organized emergence arising from interactions between diverse molecules
and cells~\cite{Emmons2016}.

The ``wiring problem'' for the {\em C. elegans} nervous system
had been posed early on with Brenner essentially raising the following 
questions:
how are the neurons spatially localized in their specific positions, 
how they connect to each other through synapses and gap junctions forming
a network with a precisely delineated connection topology, and what governs
the temporal sequence in which different neurons appear over the course
of development~\cite{Brenner1974}. Subsequent work have identified
several mechanisms underlying the guidance of specific axons 
and formation of synapses between particular 
neurons~\cite{Culotti1994,Margeta2008,Cherra2015}. However, the minutiae of the diverse
molecular processes at work may be too overwhelming for us to arrive
at a comprehensive understanding of how the complexity manifest
in the nervous system of the worm arises.
Indeed, it is not even clear that all the guidance cues that are
involved in organizing the wiring are known~\cite{Goodhill2016}.
An analogous situation had prevailed five decades earlier 
when {\em C. elegans} had been first pressed into service to understand how 
genetic mutations lead to changes in behavior of an organism. Brenner had
responded to this challenge by analyzing the system at a level 
intermediate between genes and behavior~\cite{Brenner1974}.
Thus, the problem was decomposed into trying to understand (a) the 
means by which genes specify the nervous system ({\em how is it built~?})
and (b) the way behavior is produced by the activity of the nervous 
system ({\em how does it work~?})~\cite{Brenner1974,Emmons2016}.
In a similar spirit, for a resolution of the ``wiring problem'', we may
need to view it at a level intermediate between the detailed molecular
machinery involving diffusible factors, contact mediated interactions,
growth cone guidance, etc., and the organization of the neuronal network
in the mature worm. Specifically, in this paper, we have focused on uncovering 
a set of guiding principles that appear to govern the neuronal wiring and 
spatial localization of cell bodies, and which are implemented by the
molecular mechanisms mentioned earlier (and thus genetically encoded). 
From the perspective of the
three-level framework proposed by Marr~\cite{Marr1982} for understanding
the brain~\cite{Adolphs2015}, viz., comprising (i) computational 
(or functional), (ii) algorithmic and (iii) implementation levels, 
such principles can be viewed as {\em algorithms}
for achieving specific network designs realized over the course of development~\cite{Hassan2015}.

For this purpose, we have used the analytical framework of graph theory,
which has been successfully applied to understand various aspects of brain
structure and function, in both healthy and pathological 
conditions~\cite{Bassett2008,Bullmore2009,Park2013,Stam2014,Fornito2015,Schroter2017}.
For the specific case of the {\em C. elegans} nematode, application of
such tools has revealed the existence of 
network motifs~\cite{Reigl2004}, hierarchical 
structure~\cite{chatterjee2007},
community (or modular) organization~\cite{rkpan2010} and a rich 
club of highly connected neurons~\cite{Towlson2013}. 
Comparatively fewer studies have focused on the 
evolution of the network during development of the 
nematode nervous system that we consider here~\cite{kaiser2011,Alicea2018}.
We
have integrated information about spatial location of cells,
their lineage, time of appearance, neurite lengths and network connectivity to
understand how its developmental history constrains the design of the
somatic nervous system of {\em C. elegans}, specifically the $279$ 
connected neurons 
which control all activity of the worm except the pharyngeal movements. 
Thus, our study complements existing work that has focused
more on understanding the structural organization of the network using
efficiency and optimality criteria such as minimization of the wiring cost, 
delineated by the physical distance between neurons~\cite{ahn2006,chklovskii2006,gonzalo2007,Rivera2014,tang2015,Wang2016}.

The key questions related to development that we address here involve 
the spatial location of the cell bodies ({\em why is the neuron where it
is, relative to other neurons~?}), the temporal sequence in which the
cells appear ({\em why is it that certain neurons are born much
earlier than others~?}) and the topological arrangement of their 
inter-connections ({\em why does a neuron have the links it does~?}).
As reported in detail below, we find that these questions are related
to the existence of general principles that can be expressed in terms
of different types of homophily, the tendency of entities sharing a
certain feature to preferentially connect to each other. We discern
four different types of homophily, involving respectively, 
process or neurite length of neurons, the time of their appearance, their
lineage history and bilateral symmetry. Our results help reveal that
the ganglia, anatomically distinct bundles into which 
the neurons are clustered in the nematode, are formed of
several groups (or families) of cells, neurons within each group
being closely related.

At a higher level of network organization, we show that neurons which play a vital role in coordinating activity
spanning large distances across the network by connecting together distinct 
neuronal clusters also appear quite early in the
sequence of development. This observation (along with others, such
as linking the functional type of neurons, viz., sensory,
motor and inter, to their time of appearance) helps link 
the situation of a specific cell in the temporal hierarchy to which
all neurons belong, with its function.
We also provide an analysis of the inter-relation between functional, 
structural and developmental aspects, focusing on neurons identified
to belong to different
functional circuits, such as those associated 
with mechanosensation~\cite{sulston85,wicks95,sawin96}, 
chemosensation~\cite{troemel97}, etc. This provides us with a more
nuanced understanding of the relation between the time of appearance
of a neuron and the number of its connections.
Our results suggest that 
developmental history plays a critical role in regulating the connectivity 
and spatial localization of neurons in the {\em C. elegans} 
nervous system. In other words, development itself provides key
constraints on the system design.
In addition, the tools we employ here for revealing patterns hidden 
in the lineage and connectivity
information, including novel visual 
representations of developmental history, such as chrono-dendrograms,
provide insights into principles governing the wiring of nervous systems
that may be common across several organisms.

\section{Results}
\subsection{Homophily based on multiple cellular properties
governs neuronal inter-connectivity}
Direct contact between neurons whose cell bodies are located
relatively far apart, 
through synapses or gap junctions located on their extended processes,
plays a crucial role in reducing communication delay of signals across
the entire nervous system~\cite{kaiser2006}.
This is particularly relevant for {\em C. elegans} where the majority
of synapses occur {\em en passant} (forming at axonal swellings) 
between parallel nerve process shafts that can remain close to each
other over long distances~\cite{celegans2}.
Therefore, in order to understand the principles governing the wiring
organization of the nematode nervous system, it is appropriate to
first focus on understanding how the connectivity of neurons
is influenced by the length of their neurites.

\begin{figure*}[t]
\centerline{\includegraphics[width=\linewidth]{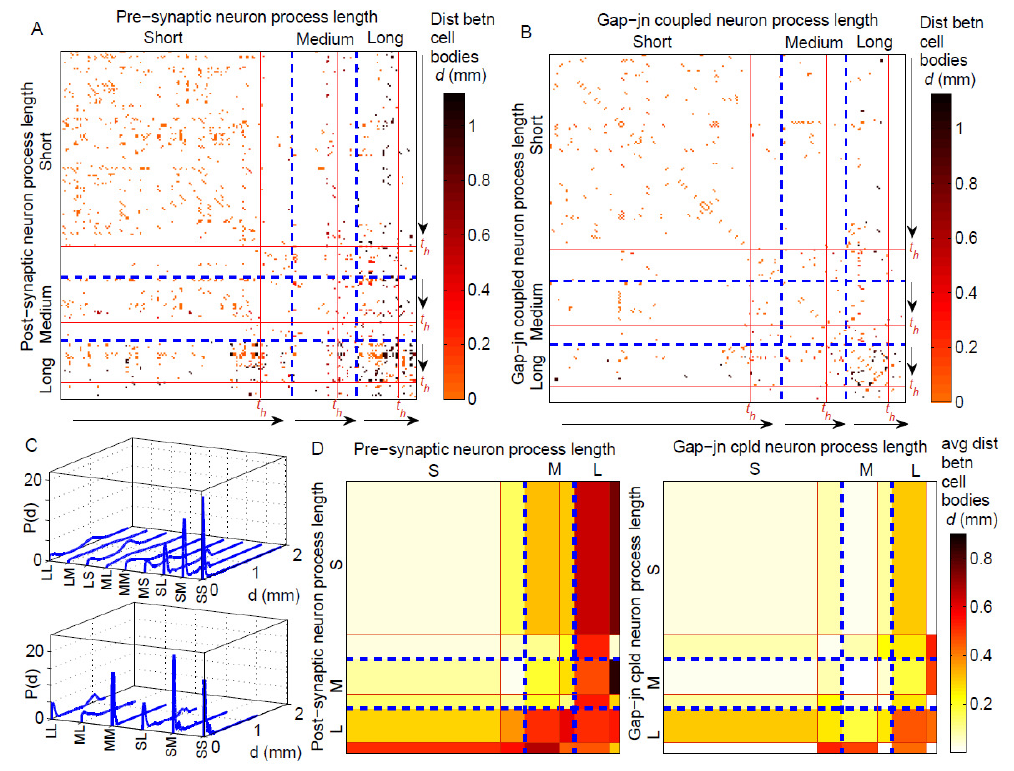}}
\caption{\textbf{Birth time cohort membership and neurite lengths
of neurons govern their connectivity.}
(A-B) Matrices representing synaptic (A) and gap-junctional 
(B) connections that exist between neurons, grouped into
three classes [indicated by blue broken lines] according to
their process lengths $\ell$ measured relative to the worm body length $L$,
viz., short ($\ell \leq L/3$), medium ($L/3 < \ell \leq 2L/3$) and long 
($\ell > 2L/3$), and ordered within each class according to birth time.
Increasing
birth time is indicated by arrows, with red lines
marked $t_h$ (time of hatching) separating neurons which differentiate 
in the embryonic stage from those born later.
Matrix entries correspond to the existence of a connection, with its color
representing the distance (measured in mm) between cell bodies of the 
corresponding neurons (see legend). We observe that there is evidence
of birth time assortative mixing, with neurons born early(later) having a higher
probability of connecting with other early(late) born neurons, which
is particularly
marked in the case of neurons having short processes.
The gap junction matrix shows a large number of entries adjacent
to the diagonal which correspond to connections between paired neurons [see
Fig.~\ref{fig5}~(A)].
(C) Distribution of distances $d$ between cell bodies of pairs of neurons
distinguished in terms of their respective process lengths (S: short,
M: medium, L: long), which are
connected by synapses (top) and gap junctions (bottom).
As synaptic connections are directed, there are nine possible
combinations of pairs of the classes (S/M/L) to which the pre- and 
post-synaptic neurons belong
(e.g., SL refers to a synapse from
a neuron with a short process to a long process length neuron).
On the other hand, as gap junctions are undirected, only six possible
combinations need be considered. 
We note the bimodal distributions of $d$ when at least one of the two
neurons connected by synapse or gap junction has a long (or medium) process.
(D) The mean distance $\langle d \rangle$ between cell bodies
of neurons connected by synapses (left) and gap junctions (right),
grouped according to their process lengths (L/M/S) [indicated
by blue broken lines] and further subdivided into those born early 
(i.e., embryonic stage) and those born late (i.e., L1, L2 or L3 stages) 
[separated by red lines]. Distances are expressed in mm (see legend for
the color code). We note that pre-synaptic neurons with long
processes tend to connect with post-synaptic neurons having 
short processes which are located far from them, corresponding to
the higher peak in the bimodal distribution for LS in top panel of (C).
Note that we have considered in this analysis the subset of 225
neurons for which information about process length is available.
}
\label{fig1}
\end{figure*}

It has also been observed that connected pairs of neurons very often 
differentiate
close to each other in time~\cite{kaiser2011}. This may suggest
that preferential connectivity among neurons according to the time of 
their birth (i.e., {\em birth cohort homophily}) is a possible basis for guiding
the network architecture. However, we need to explore the possibility
that it could be a consequence of the restrictions
on connections between neurons imposed by their respective process lengths. 
For instance, a large majority of the neurons that are born 
early, i.e., prior to hatching, are localized in the head region
and have short processes extending
to less than a third of the body length of the nematode. This could,
in principle, be sufficient
to explain the temporal closeness of connected neurons.
We have accordingly investigated the joint dependence of the
occurrence of connections
between neurons on the lengths $\ell$ of their respective 
processes, as well as, their birth times $t_b$ in Fig.~\ref{fig1}~(A-B).
The distance $d$ between the cell bodies for each pair of connected neurons
is also indicated, which makes apparent the restriction on connectivity
imposed by the process lengths. This information adds a temporal dimension
to our understanding of the organization of long-range connections 
(corresponding to high values of $d$) in the nematode nervous system.
An entry (colored point) in the $i$-th row and $j$-th column of the 
matrices shown in Fig.~\ref{fig1}~(A) and 
(B), corresponds to a chemical synapse or electrical gap junction [for 
(A) and (B), respectively] from neuron $j$ to neuron $i$ ($i,j=1, \ldots, 225$
being the indices referring to each of the neurons 
in the {\em C.~elegans} nervous
system whose process length is known).
The color represents the distance $d$ between cell bodies as per
the adjoining color bar.
The neurons (indicated along the rows and columns) are grouped according to 
their process lengths $\ell$. These are
categorized as short ($\ell \leq L/3$), medium ($L/3 < \ell \leq 2L/3$) 
and long ($\ell > 2L/3$) relative to the total body length of the worm $L$. 
Moreover, within each category, the neurons are arranged
by their time of birth in increasing order.

{\em Process length homophily.}
Even a perfunctory perusal of the two matrices makes it
apparent that the diagonal blocks in the two matrices, corresponding to 
connections between neurons having similar process length, have a 
relatively higher density of points.
This observation indicates that there is a preponderance of
connections within each group characterized by
how far their neurites extend. However, to establish that there is
indeed {\em process length homophily} which would imply, for instance, that
neurons with short processes
tend to prefer connecting to other neurons having short processes, 
we will have to compare the empirically observed number of such connected pairs
with that expected to arise by chance given
the degree (i.e., the total number of connections) of
each neuron.

To quantitatively estimate the bias that neurons may have in connecting to
neurons with similar neurite lengths, we cluster the cells into three
{\em communities} or {\em modules} which are characterized by all
their members having short, medium or long processes, respectively.
This allows us to calculate the {\em modularity} $Q$, a measure of
the extent to which like prefers connecting to like in a 
network~\cite{newman2004,newmanbook} (see Methods for details).
Low values of $Q (\sim 0)$ would indicate that there is no evidence 
to support homophily, while a relatively large positive value for
a particular module would suggest that there is a
bias for its members to preferentially connect to each other.
To see whether this is statistically significant we compare the empirical
value of $Q$ with that for an ensemble of randomized surrogates. We compute
the latter from $10^3$ networks, each constructed from the empirical
adjacency matrix (either synaptic or gap-junctional) by randomly permuting
the membership of the neurons to the three communities characterized by
the process lengths (long, medium and short) of their members.

For the entire synaptic network, we measure the empirical value of $Q$
to be $0.117$, while for the network of neurons connected by gap junctions, 
it is $0.134$. Both of these values are significantly higher than the
corresponding values for the randomized surrogates, viz., $-0.002 \pm 0.010$
for the synaptic and $-0.004 \pm 0.020$ for the gap-junctional networks, 
respectively. This suggests that neurons having similar process lengths 
do indeed have many more of their connections with each other than would
be expected simply on the basis of the number of synapses and
gap junctions possessed by each of them. Individually considering
the three communities, comprising
neurons having short, medium and long processes, respectively, also
yields $Q$ that differ significantly from the corresponding
randomized surrogates (see Supplementary Material, Table~\ref{T1}).
Thus, although the empirical values of the modularity appear to be small,
they cannot be attributed simply to fluctuations resulting from the small
numbers involved and suggests the existence of specific mechanisms that
make connections between two neurons, both of which have short (or long)
processes, more likely.

We observe a relatively high density of points in Fig.~\ref{fig1}~(A and B)
in the blocks corresponding
to cells having short processes that are born at the same epoch, i.e.,
either pre- or post-hatching. We also see this in the block in 
panel (A) corresponding to connections between pre-synaptic neurons 
with short processes
and post-synaptic neurons with medium process length. This suggests that,
apart from process length, the time of birth of the cells also determine
neuronal inter-connectivity. Indeed, earlier studies~\cite{kaiser2011} 
have shown that most of the neurons that are connected to each other
happen to be born close in time, with the probability of connection
between almost contemporaneous neurons being much more than what is
expected by chance. However, we find that the actual temporal
separation between the time of birth of different neurons does not
have any significant correlation (viz., $p\gg0.05$) 
with the probability of there being
a connection between them, either synaptic or gap-junctional.
This apparent contradiction is resolved on noting the following. 
While, within
the group of neurons born in the embryonic stage and those born
post-embryonic, there may be a great diversity in terms of birth times
(thereby significantly weakening any correlation with connection
probability), these differences are minor when viewed from the perspective
of membership in the
cohorts of those born pre- and post-hatching, respectively. 
As mentioned later, these correspond to two distinct, temporally 
separated bursts of neuronal differentiation, which provides a 
natural demarcation
of the neurons into early and late-born categories. Moreover, this {\em birth
cohort homophily} is specific to neurons whose cell bodies are located in 
close physical proximity (see Supplementary Material, Fig.~\ref{figS1}).
By comparing with randomized surrogates, we observe that connections
between neurons are not significantly enhanced if they are born in
the same epoch except for the case when the distance $d$ between their
cell bodies is short ($d < L/3$).


So far, in our consideration of how connections between neurons is affected
by their process lengths, we have not considered the information 
concerning the spatial position
of the cell bodies of the connected neurons. Consideration of this
information is important if we want to understand how activity of
spatially distant parts of the organism are coordinated through
long-range connections that allow signals to be rapidly transmitted
across relatively large physical distances.
Figs.~\ref{fig1}~(C) and (D) show how the distance $d$ between cell bodies
of connected pairs of neurons are distributed
differently according to their respective process lengths.

The top panel in Fig.~\ref{fig1}~(C) corresponds to the probability
distribution function of distance between cell bodies $d$ for neurons
connected by synapses, while the bottom panel considers gap junctions.
When both the pre- and post-synaptic neurons have short processes 
(indicated by SS in the figure),
it is expected that the cell bodies will be located close
to each other. This is indeed what is observed, with a prominent peak
of $P(d)$ occurring at extremely low values of $d$.
On the other hand, when at least one of the neurons have a long or
medium length process,
we observe that the distributions for neurons connected through synapse
are much more extended towards
higher values. For SL, LS and LL connections, we in fact observed a  
distinct bimodal character in the corresponding distribution of $d$.
This can be linked to the observation that neurons having short as well
as long processes tend to predominantly have their cell bodies
located at the head or in the tail of the worm. In contrast, neurons
whose processes are intermediate in length have cell bodies 
distributed relatively more homogeneously across the body of the organism
(see Supplementary Material, Fig.~\ref{S2}).
This can be quantified by measuring the extent to which
the cell bodies themselves are distributed along the longitudinal
axis of the nematode body in a bimodal manner
using the Bimodality Coefficient $(BC)$ metric~\cite{pfister2013}
(see Methods). A distribution is said to be prominently
bimodal if its $BC \gg BC^*$ ($=5/9$), the value of the metric for 
an uniform distribution.
We find that while the spatial positions of the cell bodies of neurons 
having short, as well as, long process are distributed in a bimodal manner
($BC_S = 0.93$ and $BC_L = 0.83$, respectively), that of
neurons with intermediate length process ($BC_M = 0.67$) are relatively
more uniformly distributed. 
Accordingly, we observe that synaptically connected pairs, in which at least 
one neuron has process of medium length, exhibit distributions of $d$ where 
bimodality is either muted (as in SM and MS) or absent (MM, ML and LM),
even though all of these distributions span a much larger range than that 
of SS.
This indicates that process length is an important determinative
factor for the occurrence of long-range connections in the nematode
nervous system.

When we consider the distribution of distances between cell bodies of 
neurons connected
by gap junctions [lower panel of Fig.~\ref{fig1}~(C)], we observe that 
connections are more likely to occur between spatially adjacent
cell bodies.
This is manifest in the distributions of $d$ being much less extended than those
seen in the case of synapses, with the exception of SL and LL which exhibit
bimodality. The distinction between the situations seen in the upper and
lower panels may arise from the fact that while synapses between two neurons
can in principle be located anywhere on their processes, gap junctions 
predominantly occur close to the cell body of at least one of the 
participating neurons.
The greater importance of relative spatial positions of neurons in
determining the occurrence of a gap junction
is manifested in terms of a stronger (anti-) correlation between
$d$ for a neuronal pair and the probability
that they are connected by a gap junction, in comparison to
a synapse (discussed later).

The detailed nature of the information about the number of neuronal pairs with 
given process 
lengths whose cell bodies are placed a specific distance $d$
apart that is provided by the distributions shown in Fig.~\ref{fig1}~(C) tends
to obscure certain gross features.
The latter can impart important insights into how process
length facilitates connections between spatially distal neurons.
Therefore, in Fig.~\ref{fig1}~(D) we display the average physical
distance between cell bodies of {\em connected} neurons which are
distinguished in terms of their process lengths (short/medium/long),
and further subdivided into those appearing in the
embryonic stage, i.e., prior to hatching (referred to as early), and those 
which appear at the post-embryonic stage (referred to as late).
For synaptic connections (shown at left),
the average $d$ for neurons with long processes (pre-synaptic) 
connected to neurons having short processes (post-synaptic) is the 
highest ($\langle d_{LS} \rangle =0.61$ mm) of all the categories considered, higher even than
that when both neurons in a connected pair have long processes
($\langle d_{LL} \rangle =0.51$ mm). Intriguingly, both of these values are 
larger than the average distance between cell bodies for
connected neurons when the pre-synaptic neurons have short processes
while the post-synaptic ones have long processes, viz, 
($\langle d_{SL} \rangle =0.32$ mm). This is consistent with
the two peaks of the bimodal distribution of $d$ corresponding to these 
connections differing substantially in amplitude - the peak 
at lower $d$ being higher for SL, while the one at higher $d$ being
larger for LS. To a lesser extent, a similar asymmetry is seen for
the average distance between connected cell bodies when one
has short process while the process of the other is of medium length
(viz., $\langle d_{MS} \rangle =0.27$ mm as compared to 
$\langle d_{SM} \rangle =0.09$ mm).

We can compare these values with the average distance between cell bodies
of {\em all} neurons, whether connected or not. For instance, the mean separation
$D$ between cell bodies of all neurons with long process lengths is 
$\langle D \rangle_{L,L} = 0.55$ mm which is almost the same as the average 
distance between every pair of neurons in which one has a short process and 
the other has a long one ($\langle D \rangle_{L,S} =0.54$ mm). 
To ensure that the difference between $\langle d_{XY} \rangle$ 
and $\langle D \rangle_{X,Y}$ (where $X,Y \in \{S,L,M\}$) is statistically 
significant, we show that it is extremely unlikely that the 
observed values of $d$ will arise by chance if random surrogates
are constructed having the same number of connected neurons as is observed empirically 
(by sampling the set of all neuronal pairs without
replacement). For instance, the $z$-score (see Methods) for
the distance between cell bodies of pre-synaptic neurons with long processes 
connected to post-synaptic neurons with short processes is $z_{LS} = 1.8$.
By contrast, considering the reverse, i.e., synapses from neurons
with short processes to those having long processes, we obtain 
$z_{SL} = -7.2$. Thus, neurons with long processes appear to form a
synapse with 
neurons having short processes whose cell bodies are located 
far away from their own much more often than that expected by chance
given the spatial positions of the cell bodies. On the other hand, 
neurons with
short processes prefer to connect to neurons with long processes
whose cell bodies are much closer to their own.
Indeed, excepting the class of LS synaptically connected
neuron pairs (i.e., pre-synaptic neuron having long process, post-synaptic 
neuron having short process) all other connected neural pair classes,
distinguished in terms of the process lengths of the two neurons,
have negative values for $z$-score (see Supplementary Material, Table~\ref{T2}).
The results indicate that the
process length of the pre-synaptic neuron is a dominant influence
deciding the average distance between cell 
bodies connected by synapses. 
It is also consistent with the possibility that a high proportion of 
synaptic contacts are occurring
close to the cell body of the post-synaptic neuron (which is closer to
the classical concept of the pre-synaptic axon connecting to a dendrite 
close to cell body of the post-synaptic neuron and not just 
making a synaptic contact anywhere on
the process). Such asymmetry between LS and SL may also have the advantage
of functional efficiency
in that the resulting connection architecture allows signals to
rapidly travel large distances across the nematode body through
long processes - thereby spreading globally using L to S connections -
and then being disseminated locally using neurons with short processes.

If we now consider the case of
neurons connected by gap-junctions [Fig.~\ref{fig1}~(D, right)],
we note that the average value of $d$ is highest for the case of cells
with long processes connecting to each other. In particular, 
unlike the situation
seen above for synaptically connected neurons, 
$\langle d_{LS} \rangle (=0.3$ mm)
is lower than $\langle d_{LL} \rangle (=0.44$ mm).
The $z$-score for the distance between cell bodies of neuron
pairs whose members belong to any of the classes S,M and L
are seen to be strongly negative, ranging between $z_{LL} = -2.2$
and $z_{SS} = -6.5$. The high statistical significance of $\langle d \rangle$ 
when
compared against the average separation between neurons $\langle D \rangle$ 
suggests
that gap junctions occur between neurons whose cell bodies lie close
to each other far more often than expected by chance (given
their positions). This is consistent with the belief that
gap junctions predominantly act to coordinate activity locally 
between neurons~\cite{xu2018}.
As alluded to above, the larger magnitude of the $z$-score values for gap junctions
as compared to synapses could possibly
indicate that these junctions tend to form much closer to the
cell body than the {\em en passant} synapses that can form
at many different locations on the extended process of a neuron.
We also note in passing another feature of gap junctional connections
between neurons which is manifest in Fig.~\ref{fig1}~(B) as a large number 
of entries in the adjacency matrix immediately neighboring the diagonal.
These correspond to a very high proportion of connections between 
bilaterally symmetric pair of neurons, e.g., AVAL and AVAR,
that is discussed later (see Fig.~\ref{fig5}). These connections may have 
the possible
functional goal of coordinating response of the nematode
nervous system to sensory inputs between the left and
right sides of the body~\cite{hall2017}.

The process length homophily between neurons that we report here 
can be attributed to
multiple possible factors. For instance, the preference of neurons having
long process for connecting to other neurons with long processes could
be an outcome of the geometry resulting from parallel fibers extending
over relatively large distances, which have a proportionately higher 
probability of forming {\em en passant} synapses with each other. 
On the other hand, the preference of neurons having short processes to
connect to each other could be tied to the fact that many of their cell 
bodies are located in close physical proximity.
This suggests an important role for the physical distance $d$ between cell
bodies in deciding connectivity between neurons. For gap junctions, we
do indeed observe a
significant correlation of $-0.66$ ($p=0.001$) between $d$
and the probability that cells are connected [consistent with the high
proportion of gap junctions occurring between neurons having
cell bodies close to each other, see Fig.~\ref{fig1}~(C), bottom].
However, for synapses, the relation
between the two is less
clear as the correlation
is not statistically significant.

Focusing only on neuron
pairs whose cell bodies are located close to each other (i.e., $d \leq L/3$ 
where $L$ is the total body length of the worm), however, we observe a very
strong correlation of $-0.94$ ($p=0.002$) between $d$ and the probability
of a synaptic connection between the two (for gap junctions, the 
correlation is $-0.91$ with $p=0.004$). This high value indicates
that synapse formation between neurons whose cell bodies are located 
near each other is indeed strongly dependent on the distance between
them. Moreover, it cannot be explained in terms of simple physical 
limits imposed by the process lengths of neurons on the farthest distance 
allowed between cell bodies of connected neurons. This is because
if we consider the correlation between $d$ and probability of connection 
only between neurons having short processes,
we obtain a value of $-0.84$ ($p=0.02$) for synapses 
and $-0.83$ ($p=0.02$) for gap junctions
(see Supplementary Material, Fig.~\ref{S3}).

A possible explanation for the weakening of the relation between
connection probability and the 
physical distance separating the cell bodies when all neurons are
considered could be because,
even though neurons born in    
close physical proximity have a higher probability of getting connected, 
it is masked by the cells moving apart subsequently over the course
of development.
In the absence of information about the location of the cell bodies
at the time synaptogenesis happens, we can probe this indirectly by
considering how the probability of connection between two cells depends 
on how closely they are related in terms of lineage - as cells having
common ancestry also tend to be born adjacent to each other.

\begin{figure*}[t]
\centerline{\includegraphics[width=\linewidth]{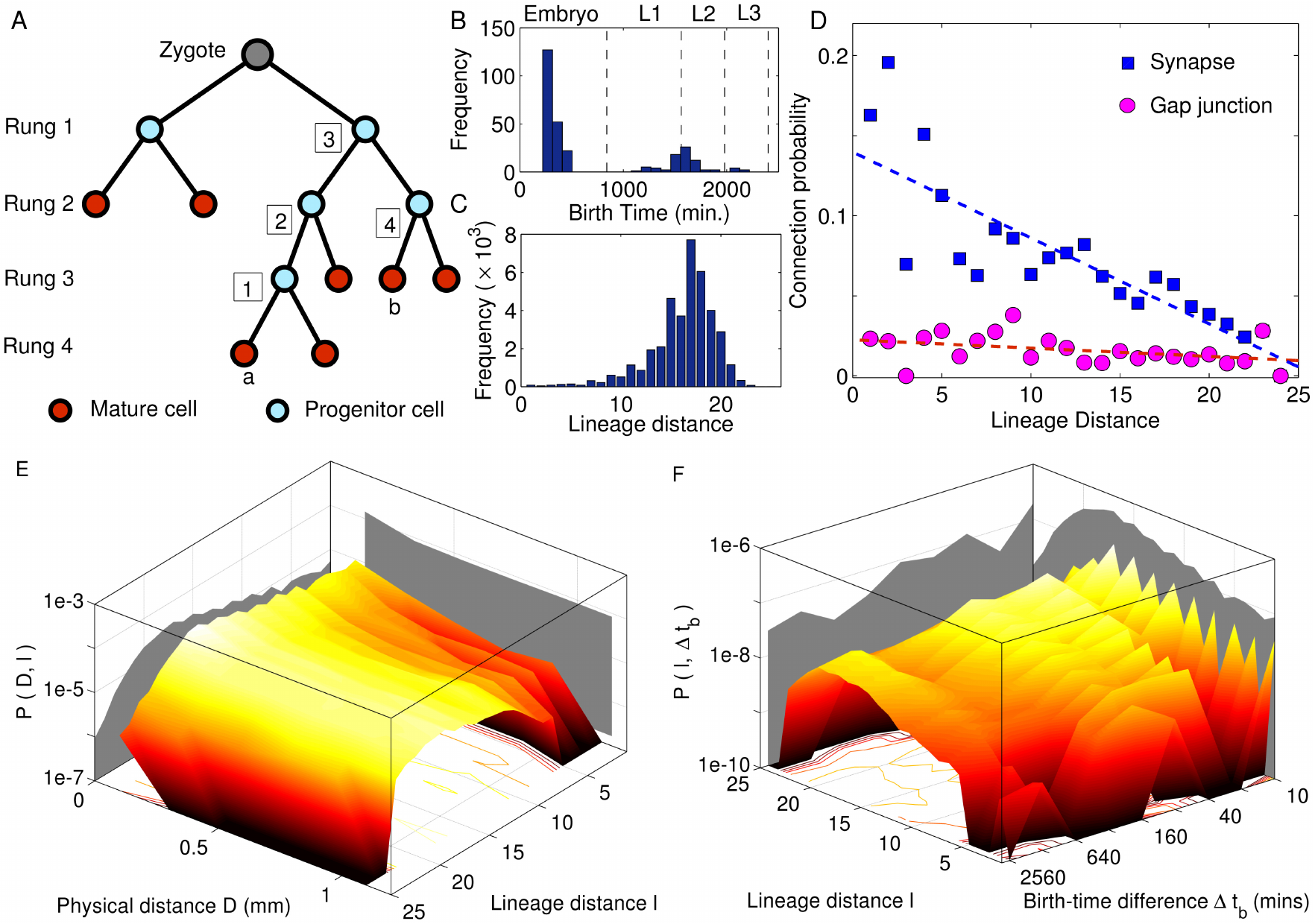}}
\caption{\textbf{Lineage of neurons affects their synaptic connectivity
and spatial localization.} 
(A) Schematic diagram of a lineage tree of cells resulting from 
consecutive mitotic divisions of the zygote. The terminal nodes
of the tree correspond to terminally differentiated mature cells
(shown in red) while other nodes represent progenitors (shown in blue) 
that appear at different rounds of cell division.
Cells born at each round of cell division are indicated by the
corresponding rung of the tree they belong to, the numerical value
for the rung (shown at the left) being the number of divisions
starting from the zygote.
The lineage distance $l$ between a pair of mature cells is measured as the
total number of cell divisions leading to each from their 
common progenitor. An example of lineage distance measurement 
is shown in the figure for the pair of cells $a$ and $b$ which
are separated by
four cell divisions (the distance of $a$ from each of the intermediate 
dividing progenitors is indicated in the figure).
(B-C) Frequency distributions of the birth time of different neurons
(B, separated into the different developmental stages) and
the lineage distances for each pair of neurons (C).
(D) The probability of a pair of neurons to be connected through a
synapse decreases with increasing lineage distance between them,
as indicated by a statistically significant linear correlation
between the two ($r = -0.86$, $p<10^{-7}$). In contrast, there is 
no significant variation of the probability of gap junctional connection
with lineage distance.
(E-F) Joint probability distributions of lineage distance $l$ along with
distance between cell bodies $D$ (E) and
birth time difference $\Delta t_b$ (F)
between all pairs of neurons.
The marginal distributions for the corresponding quantities are 
indicated in the bounding surfaces.
Contours for the distributions are indicated at the base of each
figure.
We notice that the distribution of physical distances in (E) exhibit
a bimodal nature. However, cells which are closely related 
in terms of lineage ($l<5$) also has a high probability of
being physically located nearby (indicated by a prominent
peak at the lower end of the distribution of $D$) which suggests
that lineage influences spatial localization of cells. 
In panel (F), the distribution shows peaks at odd values of the lineage
distance 
(particularly for low $\Delta t_b$) suggesting that neurons
born close in time are located at the same rung on the lineage tree. 
}
\label{fig2}
\end{figure*}

{\em Lineage relation between neurons constrains the distance between
their cell bodies, as well as, the likelihood of a synaptic connection 
between them.}
Cell lineage provides knowledge of the developmental trajectory in
all metazoa, being 
defined by successive divisions starting from the zygote to the final 
differentiated cell. In most animals, the identity of any terminal
node of the lineage tree, known as
cell fate, is determined by intrinsic and extrinsic factors, as well as, 
interactions with neighboring cells. This introduces sufficient
variability in the developmental path so as to make lineage relationships 
discernible only at the level of cell groups rather than individual 
cells~\cite{wood88}.
However, some organisms such
as nematodes exhibit
an almost invariant pattern of somatic cell divisions that is identical across 
individuals, and in the case of {\em Caenorhabditis elegans}, is known
in its entirety~\cite{sulston77,sulston83}.
Thus, the lineage tree of the organism provides us with a complete fate map
at single-cell resolution~\cite{giurumescu2011}.
The schematic representation of such a tree shown in Fig.~\ref{fig2}~(A)
depicts successive mitotic cell divisions starting from a
zygote that, through intermediate progenitor cells, eventually 
differentiate into mature neuronal cells.
Each successive cell division (beginning from the zygote) corresponds 
to different rungs in the tree used to label the resulting daughter cells.
The difference between any two cells in terms of their lineage can thus 
be quantified by their lineage distance, i.e., their separation on the tree 
measured as the total number of cell divisions that leads to each of them
from their last common progenitor.

Apart from the lineage tree, crucial information on the
relationships between different cells that stem from their developmental
history is provided by the knowledge of birth times of the
individual mature neurons, i.e., the specific instant 
in developmental
chronology of the nervous system at which each neuron differentiates. 
Fig.~\ref{fig2}~(B) shows the distribution of birth times for all cells 
belonging to the somatic nervous system of {\em C. elegans}, indicating 
that development of the system occurs in two bursts clearly
separated in time. The `early burst', during which the bulk, viz., 72\%, 
of the neurons are born,
occurs at the embryonic stage of
development, while the more temporally extended `late burst' spans across
the L1 and L2 stages.
This information, in conjunction with a simple generative model for 
reconstructing the
lineage tree through successive cell divisions,
can be used to explain the distribution of lineage distance shown in 
Fig.~\ref{fig2}~(C).
As at each node of the lineage tree a cell divides into at most two daughter cells, we can view
it - at least in the first few rungs belonging to the early proliferative phase - as a balanced binary tree, with the
number of cells that appear in each rung $R$ increasing exponentially with $R$ (upto $R=10$ in {\em C. elegans}, see
Supplementary Material, Fig.~\ref{S4}). Within
the AB sublineage of cells to which almost all the neurons belong, 
the maximum lineage distance that can occur between two cells which
are placed in 
rungs $R_1$ and $R_2$, respectively, is given by 
$l_{max} (R_1,R_2) = (R_1 - 1) + (R_2 - 1) - 1.$ 
Thus, the distribution of lineage
distances has an exponential profile upto $l=17$. Beyond rung $10$, the subsequent branching of the nodes in the binary tree
reduce markedly as many of the divisions terminate in differentiated neurons (and occasionally programmed cell death) or 
lead to non-neuronal fates (so that their further divisions are not considered for the purpose of
this study).  This can be seen to result in the lineage distance 
distribution {\em decreasing} exponentially for $l>17$, with a 
maximum lineage distance of $25$.
A more detailed theoretical model of the lineage relationships between 
neurons resulting from their developmental history
can be constructed as an asymmetric stochastic branching process 
(see Methods). Here, beginning with a single node that corresponds to the zygote, at each iteration every node that appeared 
during the preceding iteration is considered in turn for giving rise to each of two possible branches with probabilities
$P1$ and $P2$ ($P1 \geq P2$) that result in further nodes. 
By considering the actual lineage tree, these asymmetric branching probabilities in the model
were fixed as
$P1=1$ and $P2=0.85$ until rung 9 and for later rungs they were set to $P1=0.25$ and $P2=0.2$. 
For these values of $P1$ and $P2$, the trees generated by the model exhibited properties that were statistically similar to the empirical 
lineage tree (see Supplementary Material, Fig.~\ref{S4}).


Going back to the question we had posed earlier, viz., how does the 
lineage distance $l$ between cells affect the probability
that they are connected by synapses, we observe from Fig.~\ref{fig2}~(D)
that there is indeed a strong correlation of $-0.86$ ($p<10^{-7}$)
between the two. This observation provides evidence of {\em lineage
homophily} being one of the key principles governing connectivity
of the nematode nervous system.
However, for gap junctions we do not see any significant relation
between the probability of a connection between two neurons
and how close they are in terms of their ancestry.
These observations suggest the following plausible scenario, viz.,
synaptogenesis can occur early, just after neurons are born,
while gap junctions are established much later during development,
when neurons have more or less moved to their 
final positions. Thus, changes in the locations of cell bodies from
that they occupied initially (i.e., at the time the corresponding 
neurons differentiated) which are brought about by the
appearance of cells born later through 
subsequent cell-divisions, result in a weak correlation
between synaptic connection probability and physical distance
separating the cell bodies as alluded to earlier. It may also lead to 
neurons of dissimilar lineage (whose cell
bodies need not initially be close) eventually move in physical
proximity of each other allowing the possible formation of gap junctions
between them.

The connection between lineage distance $l$ and physical distance $D$ between
cell bodies of neurons (whether connected or not), which has been mentioned
earlier, is illustrated by the joint probability distribution $P(D,l)$ shown
in Fig.~\ref{fig2}~(E). 
In particular, cells having short lineage distance, viz., $l \leq 5$,
tend to have their cell bodies located close to each other, as indicated
by the function being peaked towards lower values of $D$. However,
cells that are farther apart in terms of lineage can occur at different
distances from each other, resulting in the overall bimodal form for
the marginal distribution of $D$.
A similar nuanced relation between lineage distance for two neurons
and the difference of the times $\Delta t_b$ in which they are born is 
indicated by the joint probability distribution $P(l,\Delta t_b)$ shown
in Fig.~\ref{fig2}~(F). We note that for small $l$ ($l \leq 5$), the
distribution peaks at low values of $\Delta t_b$ indicating that closely
related neurons tend to be born within a short time interval of each other. 
We also observe that the distribution of $l$ between neurons that 
differentiate at around the same time (i.e., for low $\Delta t_b$)
tends to alternate between peaks and troughs for odd and even values,
respectively. This is easy to explain if neurons that are contemporaneous
occur at the same rung (as, by definition, neurons at the same rung will
have odd values of lineage distance between themselves).

\begin{figure*}[t]
\centerline{\includegraphics[width=\linewidth]{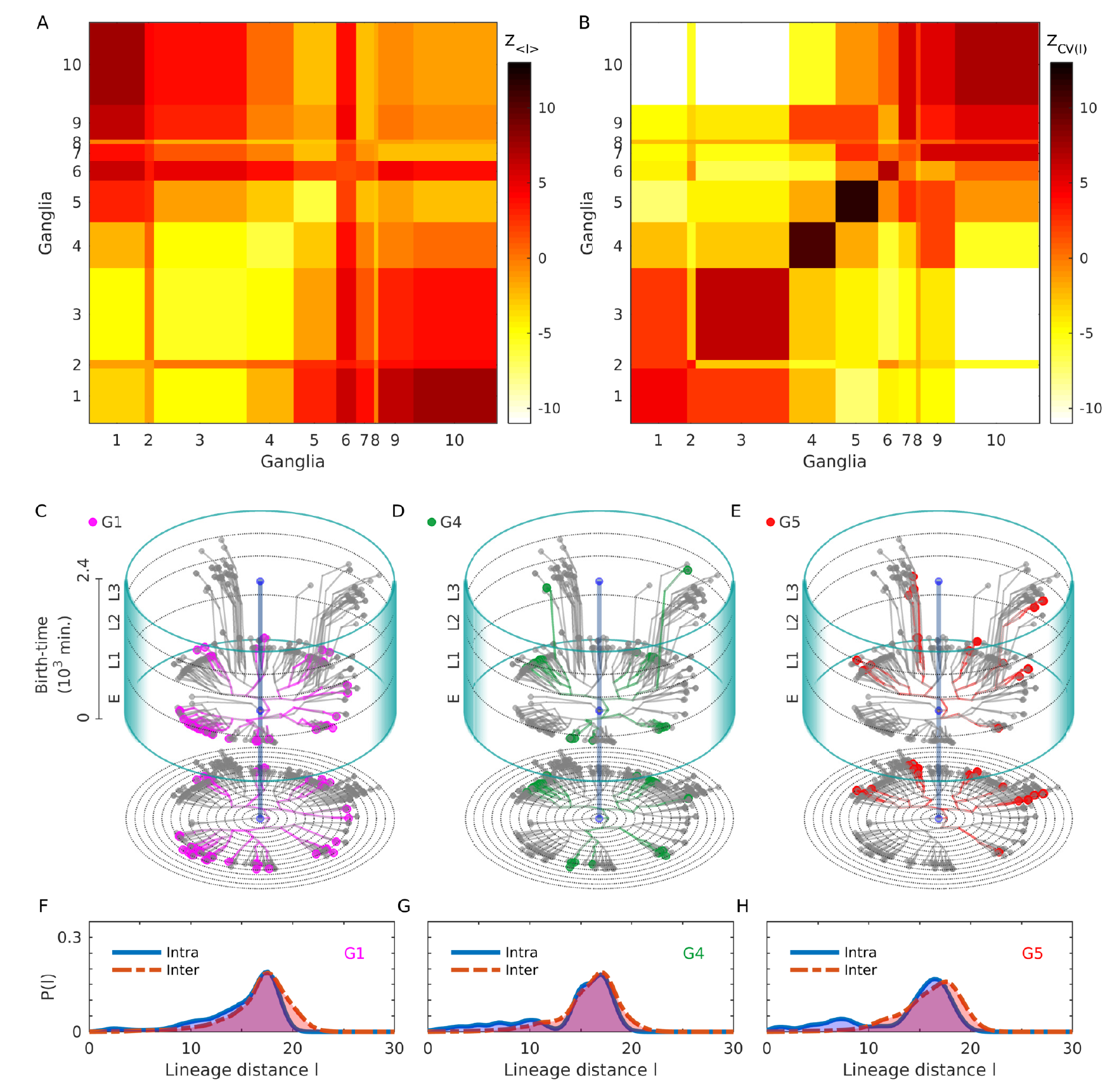}}
\caption{\textbf{Lineage distance reveals developmental patterns of ganglia.}  
(A-B) Statistically significant features of the distribution of  
intra and inter-ganglionic lineage distances,
quantified by deviations of the mean $\langle l \rangle$ (A) and coefficient of variation $CV$
(B), from a surrogate ensemble
of randomized lineage trees of neurons in the C. elegans somatic nervous system.
These deviations (measured by $z$-score) 
show that the mean intra-ganglionic lineage distances 
(represented by diagonal blocks of the matrix)
are significantly lower than that of the 
inter-ganglionic lineage distances (off-diagonal blocks). 
By contrast, CV for the 
intra-ganglionic lineage distances are significantly higher 
than that of the inter-ganglionic lineage distances.
(C-E) Developmental 
chrono-dendrograms for three representative ganglia (viz., G1, G4 and G5) show that each 
comprises multiple localized clusters of neurons occurring at different  locations on the
developmental lineage tree, explaining the statistically significant
deviations of the mean and CV for intra-ganglionic
lineage distances.
Colored nodes represent neurons belonging to the specified ganglion
while grey nodes show the other neurons. 
Branching lines trace all cell divisions starting from the 
single cell zygote (located at the origin) and terminating at each differentiated neuron. 
The time and rung of 
each cell division is indicated 
by its position along the vertical and radial axis respectively.    
The entire time period is 
divided into four stages, viz., Embryo (indicated as E), L1, L2 and L3. 
A planar projection at the base of each cylinder shows 
the rung (concentric circles)
of each progenitor cell and differentiated neuron.
(F-H) The probability distribution 
functions for the intra-ganglionic lineage distances show bimodality (unlike that of the
inter-ganglionic distances), which is consistent with the segregation of a ganglion
into multiple clusters along the chrono-dendrogram.
The different ganglia are indicated by symbols
G1-G9 (1: Anterior, 2: Dorsal, 3: Lateral, 4: Ventral, 
5: Retrovesicular, 6: Posterolateral, 
7: Preanal, 8: Dorsorectal and 
9: Lumbar) and the Ventral cord as G10.}
\label{fig3}
\end{figure*}

{\em The different ganglia comprise clusters of closely related neurons.}
The compelling association between lineage and physical proximity
of neurons alluded to above is manifest in the spatial organization
of the cell body locations. It is particularly
conspicuous in the clustering of neurons into anatomically 
distinct bundles that are referred to as ganglia.
These structures, characteristic of nematode nervous systems, contain
only cell bodies of the neurons with their axonal and dendritic processes
located outside of the bundles~\cite{Schafer2016}. The somatic nervous 
system comprises nine such spatially localized clusters, viz., 
anterior, dorsal, lateral, ventral, retrovesicular, posterolateral, 
preanal, dorsorectal and lumbar ganglia, with the remainder belonging to
the ventral cord. 
Comparison of the distributions of intra-ganglionic lineage distances 
(i.e., between pairs of neurons located in the same ganglion) with that
of inter-ganglionic lineage distances (i.e., between neurons
in different ganglia) provides an insight into how these bundles
can be interpreted from a developmental perspective.

We first note that the mean of the lineage distances 
$\langle l \rangle$ within a given ganglia 
are typically much smaller than those between different ganglia. 
Moreover, as seen from Fig.~\ref{fig3}~(A), the mean of the 
intra-ganglionic lineage distances for most ganglia are 
significantly small, which we determine by comparing with values of 
$\langle l \rangle$ obtained from ensembles of $10^3$ surrogate lineage
trees where the identity of each of the leaf nodes (i.e., the 
differentiated neurons) has been
randomly permuted. 
This randomization decouples the ganglionic membership of the neurons
from their position on the lineage tree while keeping the lineage
distances between cells invariant, consistent with our null hypothesis
that the ganglion to which a neuron belongs is
independent of its developmental history.
The observed mean intra-ganglionic lineage distances deviate markedly from
those obtained from the surrogate trees (as measured by $z$-score, see
Methods), indicating that
neurons in a ganglion are much more closely
related to each other than expected by chance. 

However, when we consider the coefficient of variation ($CV$), a relative
measure of the dispersion in the lineage distances within a ganglion 
or between two ganglia, we note that this is almost always greater
for intra-ganglionic, compared to the inter-ganglionic,
lineage distances [Fig.~\ref{fig3}~(B)]. We can again establish the
statistical significance by measuring the same
quantities for the ensemble of surrogate lineage trees mentioned above
and quantifying the difference between the actual tree and the randomized
ensemble using $z$-scores. The large values of $z$ for CV in most of the
diagonal blocks (corresponding to intra-ganglion dispersion) shown
in Fig.~\ref{fig3}~(B), suggests
that the relatedness between neurons in a
ganglion shows a much larger variability than expected by chance.

The apparent contradiction between the results mentioned above,
viz., that a majority of the neurons in a ganglion 
have a shared lineage
while, at the same time, exhibit a high degree of diversity in their 
lineage relations, is easily resolved on inspecting the chrono-dendrograms
that visually represent the complete developmental
trajectory for each of the ganglia [shown in Fig.~\ref{fig3}~(C-E), for the anterior,
ventral and retrovesicular ganglia; see Supplementary Material, 
Figs.~\ref{S5}-\ref{S7} for the others]. 
While the lineage tree shown in each of these figures is, of course, 
identical, the neurons that belong
to a particular ganglion are distinguished (by color) in the corresponding
chrono-dendrogram, allowing us to note at a glance how all the members 
of the given ganglion relate to each other.
We note that the differentiated neurons that constitute a ganglion 
are typically organized into multiple clusters, each of which are
highly localized on the lineage tree. In other words, a ganglion
comprises several `families' of neurons emanating from different
branches of the tree, with each family composed of 
closely related cells sharing a last common ancestor separated
from them by only a few cell divisions.

The grouping of the cells belonging to a particular ganglion
into distinct clusters, which are widely separated on the lineage tree,
is reflected in the bimodal nature of the distribution of intra-ganglionic
lineage distances [Fig.~\ref{fig3} (F-H)]. In contrast to the unimodal
distribution seen for inter-ganglionic lineage distances, the neurons
within a ganglion could either have (i) extremely low distances to
cells which belong to their own `family' or (ii) large distances to
cells belonging to the other `families' that constitute the ganglion.
These manifest, respectively, as a smaller peak at lower values and a larger
peak at higher values of $l$ seen in Fig.~\ref{fig3} (F-H). The
bimodality gives
rise to a large dispersion and hence a value for the CV of lineage
distances that is higher than expected. 
Note that the peak at higher $l$ for this distribution
almost coincides with the peak of the inter-ganglionic $l$ distribution,
which is expected as the latter is dominated by cells that are not
closely related. Thus, the presence of the second peak at lower values
of $l$ in the intra-ganglionic
distribution reduces the mean lineage distance
for cells within a ganglion, compared to that for cells belonging to
different ganglia. Conversely, the absence of multiple peaks in
the inter-ganglionic distribution provides for a smaller value of the CV
compared to the case for the intra-ganglionic distribution.
Thus, these results explain the apparently contradictory
coexistence of low mean value and high CV for lineage distances of
neurons within a ganglion, which is related to the localization
of the developmental trajectories of cells belonging to it into
distinct groups visible in the lineage tree.
This clearly demonstrates that the spatial segregation of
neurons into ganglia is shaped by the relations between
the constituent cells which arise from their shared developmental history. 

\begin{figure*}[t]
\centerline{\includegraphics[width=\linewidth]{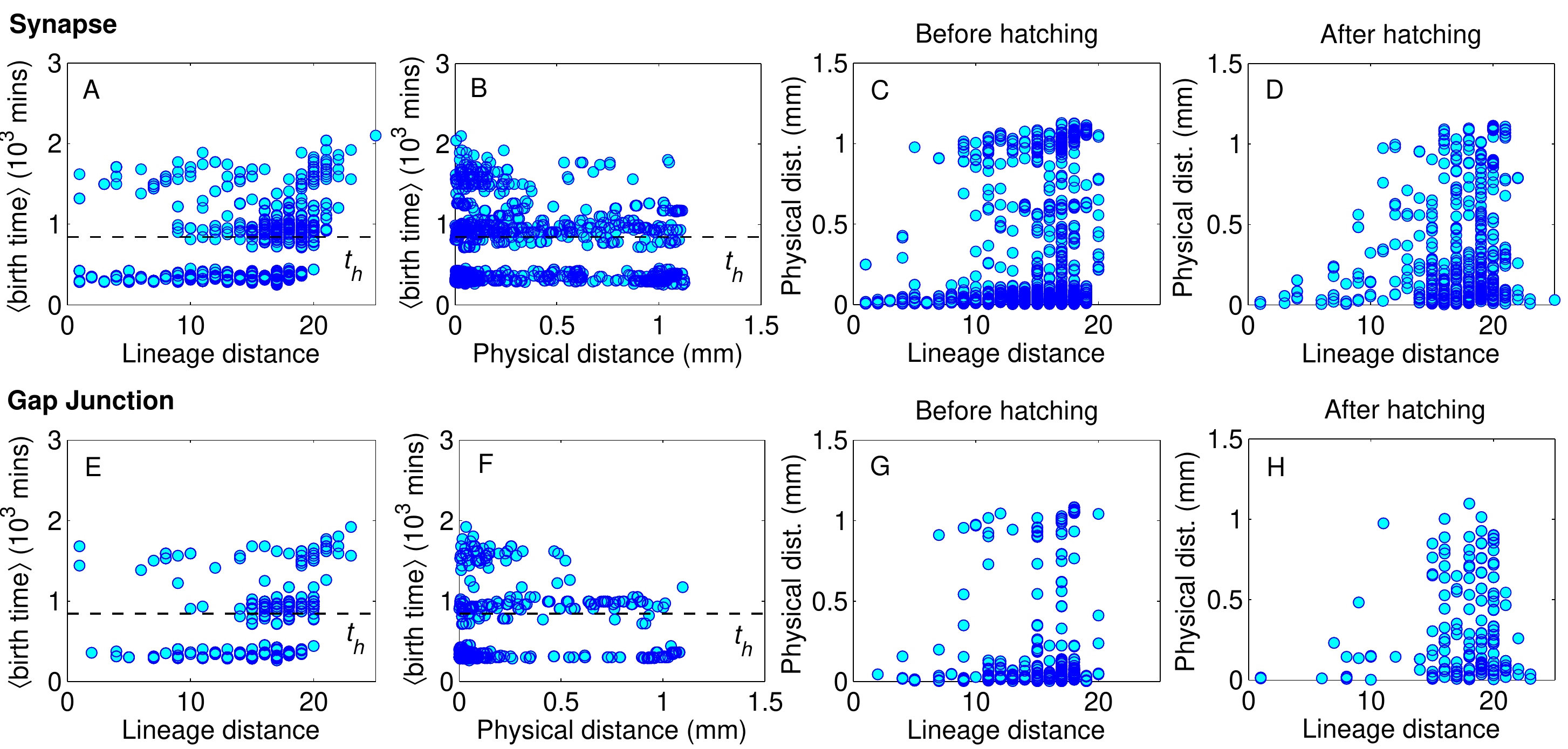}}
\caption{\textbf{Birth times and lineage distances constrain connections
between neurons whose cell bodies are spatially distant from each other.}
(A-B) The mean birth time of synaptically connected pairs of neurons
exhibit a trimodal distribution, with connections clustering into
three temporal groups corresponding to those (i) between neurons that
are both born early, i.e., in the embryonic stage, (ii) between one
born early and the other born late (i.e., in the post-embryonic
stage), and (iii) between neurons that are both born late.
The hatching time $h_t$ separating the embryonic from other developmental
stages is indicated by the broken line.
We note from panel (B) that when both neurons are born late (corresponding to
the uppermost cluster of connections), synaptic connections
are more likely to occur between neurons whose cell bodies are located
close to each other.
(C-D) Synaptic connections between
neurons that are closely related to each other in terms of lineage ($l<10$)
occur almost always when their cell bodies are in proximity, regardless
of the time of birth of the neurons. We note that this restriction 
is more pronounced than observed
in Fig.~\ref{fig2} (E), where $P(D,l)$ shows a prominent peak at
the lower end of $D$ for small $l$ suggesting that most closely related 
neurons (whether connected or not) typically have short distances 
between their cell bodies.
(E-H) Neurons connected by gap junctions show patterns similar to those seen 
in the case of synaptic connections.
}
\label{fig4}
\end{figure*}

{\em Connected neurons.} 
Having considered the distribution of physical distance, lineage distance
and birth-time differences between all neuronal pairs in the somatic nervous
system, we now focus on the subset of connected pairs to see
how the above factors may constrain the probability that a neuron 
has a direct interaction with another. Fig.~\ref{fig4} shows the
inter-relations between similarity of ancestry, spatial separation and
birth times for each pair of neurons that are linked
either by synapses (top row) or gap junctions (bottom row).
The clustering of mean birth times of the connected pairs into three
distinct groups (seen in panels A-B and E-F) is a consequence 
of the two bursts of neuronal 
differentiation widely separated in time [seen in Fig.~\ref{fig2}~(B)].
Thus, the lower and upper clusters correspond to connected neurons
both of which appear in the course of the same developmental burst
(early and late, respectively), while connections between
neurons that arose during different bursts populate the intermediate cluster.

In Fig.~\ref{fig2}~(F) we had already seen that closely related neurons 
tend to have similar birth times. This helps explain why, as seen in
Fig.~\ref{fig4}~(A), whenever synaptically connected neurons have short lineage
distance to each other, they also happen to belong to the same
developmental burst epoch. 
However, apart from the relative differences in the birth times, the
actual time of differentiation also determines the occurrence of
a synapse between neurons. Indeed, it is known from Ref.~\cite{kaiser2011}
that about 68\% of long-range synaptic connections occur between neurons 
both of which are born
in the early burst of neuronal differentiation.
This is complemented by Fig.~\ref{fig4}~(B) which shows that 
synapses between neurons, whose cell bodies are separated by 
large distances, mostly occur when
at least one of the neurons was born early. 
Conversely, when both neurons are born in
the late burst, such long-range links become extremely unlikely.
Indeed, the distribution of distances between cell bodies of
connected neurons (see Supplementary Material, 
Fig.~\ref{S8}, that compares the
empirical data with degree-preserved randomized networks where the
connections are made according to constraints imposed by the length
of processes of each neuron) show that long-range
connections in the nematode typically do not occur significantly more often
than that expected by chance, given the process lengths of the neurons.
Thus, specific mechanisms for explaining the occurrence of such 
connections maybe unnecessary given that {\em en passant} synaptic contacts
form between neighboring parallel neuronal processes.
In contrast, short range connections are much more numerous than that
seen in the random surrogate networks. This suggests that active processes may
be driving synaptogenesis~\cite{Margeta2008,Cherra2015} between neurons lying in close proximity, for
example, chemoattractant diffusion~\cite{TessierLavigne96,Chen2009,Kolodkin2011}.
Furthermore, the exceptional feature of early pre-synaptic
neurons having long-range connections
to late post-synaptic neurons much more often than is expected by chance
could suggest a possible role of 
fasciculation in this process~\cite{Kaiser2017}. 
For instance,
late-born neurons could be following the extended processes of earlier neurons
to connect to cell bodies placed far away.

In Fig.~\ref{fig4}~(C) and (D) we compare explicitly the pre- and
post-hatching scenarios in order to see whether early and late-born
neurons differ in terms of how the synaptic connections between them
are influenced by the lineage and/or physical distances between them. 
We note that for both groups of cells, closely related neurons that 
are connected by synapse also
happen to occur at spatially proximate locations. This is consistent
with Fig.~\ref{fig2}~(E) where the peak in the joint probability
distribution of all neuronal pairs with lineage distance $l$ and
physical distance $D$ is observed to occur at low $D$ when $l$ is small.
Qualitatively similar results are observed when we consider neuronal
pairs connected by gap junctions [see panels E-H of Fig.~\ref{fig4}].

The results reported above provide remarkable evidence for the role that 
developmental attributes (viz., lineage distances and birth-times
of neurons) play in shaping the spatial organization of cell bodies and the 
topological structure of 
the connections in the
somatic nervous system of the worm.
However, the process length homophily described earlier appears to
be independent and cannot be explained as a consequence of lineage 
homophily. The chrono-dendrograms (see Supplementary Material, Fig.~\ref{S9}) 
showing the positions of neurons with short, medium and long processes, 
respectively, on the lineage tree
indicate that neurons having a particular process length do not cluster
together.
This suggests that neurons with extremely similar lineage may have very
different process lengths (and vice versa), so that the observed
bias in the connection probability between neurons having processes of
similar length cannot simply be attributed to a common lineage.

{\em Bilaterally symmetric neurons.}
The major fraction ($\approx 66\%$) of neurons belonging to the somatic 
nervous system of {\em C. elegans} occur in pairs. These are located along the
left and right sides of the body of the nematode in a bilaterally symmetric
fashion. While there are instances of bilaterally symmetric neurons exhibiting 
functional lateralization (e.g., ASEL/R, see Ref.~\cite{hobert2014genesis}),
the vast majority of the left/right members of such pairs remain in the 
symmetrical ``ground state'', i.e., they are
indistinguishable functionally, as well as, in terms of anatomical features
and gene expression~\cite{hobertWormBook}. In particular, whenever
one member of a bilaterally symmetric pair occurs in any of the known functional
circuits, 
the other also appears in it without
exception. While it is known that 
this symmetric nature is manifested 
in the spatial arrangement (e.g., location of the cell bodies) and 
connection structure of paired neurons, 
here we ask whether bilaterally symmetric
neurons share a similar network neighborhood, i.e., whether there is
a high degree of overlap between the neurons that each of them
connect to, or indeed whether they have a significantly higher
probability of being connected to each other. The latter assumes
importance in view of the fact that it is the direct contact between
the paired cells AWCL/R that trigger asymmetrical gene expression resulting
in differential expression of olfactory-type G-protein coupled receptors
in the neurons~\cite{hobertjohnston2002}.

\begin{figure*}[t]
\centerline{\includegraphics[width=\linewidth]{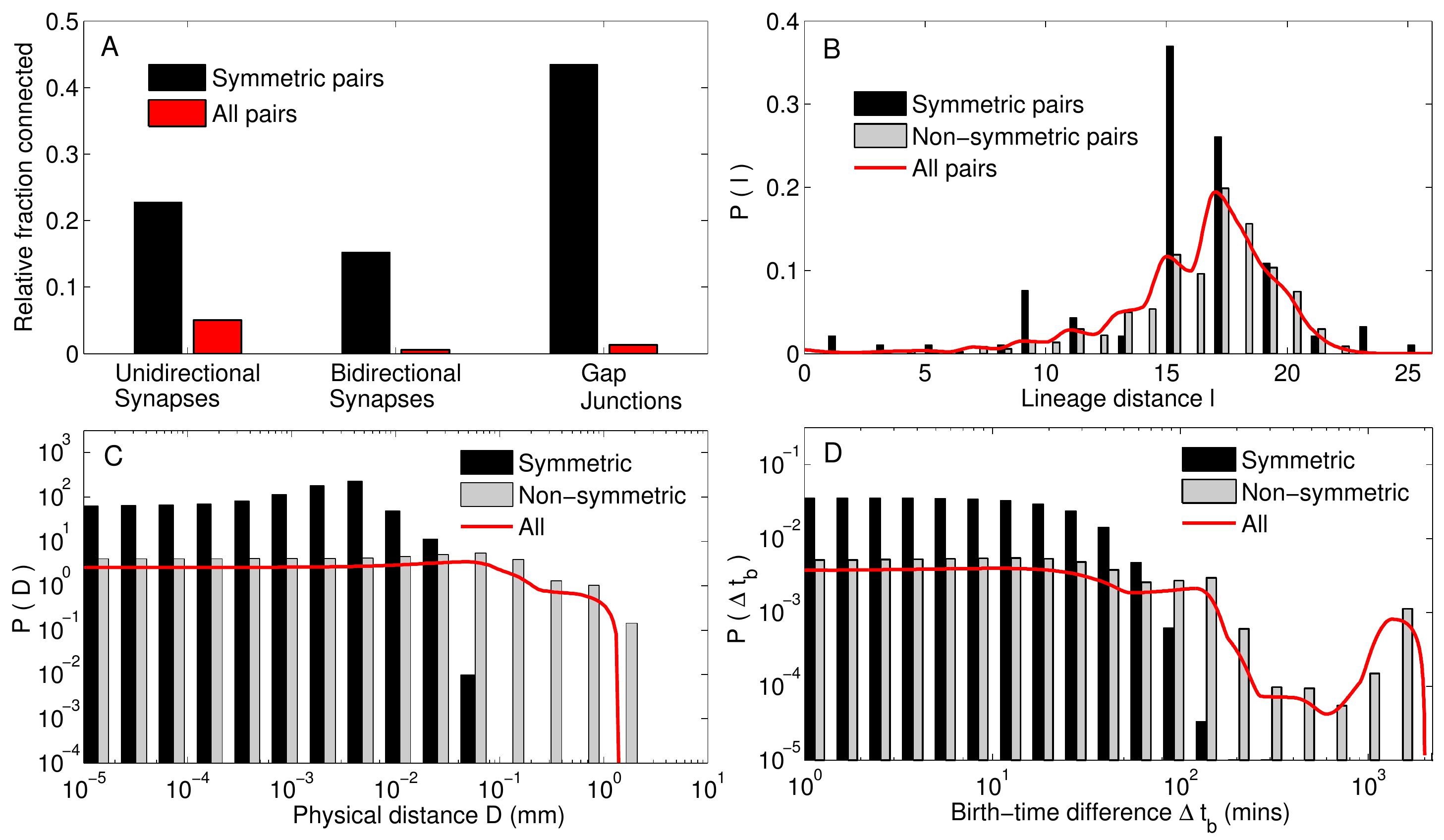}}
\caption{\textbf{Symmetrically 
paired neurons have a high probability of being connected
and also exhibit strong association in 
their birth times and spatial positions.}
(A) Bilaterally symmetric neurons that are positioned on the left and right
of the body axis of the organism tend to have a much higher probability
of synaptic, as well as, gap junctional connections between them, compared
to that for all pairs of neurons. In addition, the synapses are highly
likely to be reciprocal (bidirectional). 
(B) The distribution of lineage distances between paired neurons show
that the mean value is lower than that for all neurons. We note that
almost all lineage distances between symmetric neurons are odd-valued
suggesting that they occur at the same rung of the lineage tree.
(C-D) Symmetrically paired neurons have cell bodies located in physical
proximity of each other (C)
and are born close in time as indicated by low birth-time
differences $\Delta t_b$ (D), compared to all pairs of neurons.
}
\label{fig5}
\end{figure*}

Fig.~\ref{fig5}~(A) shows that indeed the left/right members of a symmetric 
pair have a much higher probability of connection between them than
any two arbitrarily chosen neurons belonging to the somatic system.
Moreover, $15\%$ of the bilaterally symmetric pairs have reciprocal 
synaptic connections with each other, compared to less than $1\%$
of all neuronal pairs being connected in such a bidirectional manner.
We can further distinguish the symmetric neuron pairs into those which 
originate from an early division across left/right axis of the common ABp 
blastomere (i.e., they have similar lineage differing
only in the early cell division event ABpl/r) and those where members of a pair 
originate from non-symmetric blastomeres 
(e.g., ABal and ABpr)~\cite{hobert2014genesis}.
These two distinct origins of the bilaterally symmetric
neurons are reflected in the two peaks of the distribution
of lineage distance between the left/right members of each pair
seen in Fig.~\ref{fig5}~(B), with only the latter category of paired neurons
that do not share a bilaterally symmetric lineage history 
having low values of $l$.
The synaptic connection probability between the members of pairs belonging
to these two classes differ only by a small amount ($0.26$ for the former
and $0.19$ for the latter, with the corresponding numbers reducing
to 0.18 and 0.12, respectively, when we consider reciprocal synapses).
The occurrence of gap junctions between bilaterally symmetric neurons
is seen to be exceptionally high ($43\%$ of such pairs being connected) 
compared to that for the entire system,
with no distinction in numbers being observed between the two categories
of symmetric pairs.
This preponderance of gap-junctional connections between bilaterally 
symmetric neurons (also indicated by the band diagonal structure of the 
connectivity matrix shown in panel (B) of Fig.~\ref{fig1}) 
suggests that their activity is
highly coordinated. This may possibly explain the co-occurrence of
both members of a symmetric pair in the different functional 
circuits.

In addition to exhibiting a high probability of being connected directly,
bilaterally symmetric neurons are also characterized by a high degree
of neighborhood similarity. 
Fig.~\ref{S10} 
in Supplementary Material shows the magnitude of overlap between the neurons
that each member of a pair is connected by a synapse (either pre- or
post-synaptically) or a gap junction, which is seen to be much higher than
that for any two arbitrarily chosen neurons.
This is consistent with the left/right neurons in the majority of 
bilaterally symmetric pairs having an identical
role in terms of the mesoscopic organization of the network 
(see discussion related to Fig.~\ref{fig7}).
The large number of neighbors that
paired neurons share in common is a striking feature that cannot
be explained from their physical proximity alone.

We note that
almost all lineage distances between symmetric neurons are odd-valued
suggesting that they are born at the same rung of the lineage tree.
The only exception is the pair AVFL/R, whose members have distinct
non-symmetric lineage history, with a lineage distance of 8.
Given their shared lineage, it is perhaps unsurprising that 
most bilaterally symmetric paired neurons
also exhibit strong associations in their physical locations and
birth times.
Panels (C-D) show that a large fraction of the left/right members 
have cell bodies  that are located in close physical
proximity of each other (C)
and are also born close in time as indicated by low birth-time
differences $\Delta t_b$ (D), compared to all pairs of somatic neurons.
Indeed we note that the only exception is the late-born pair SDQL/R with
bilaterally symmetric history
whose members are located in the anterior and posterior (respectively) parts
of the organism, the physical distance between the cell bodies being $0.5$ mm. 

\subsection{Temporal hierarchy of the appearance of neurons during
development is associated with their functional identity}
We have been focusing, so far, on the various properties related to
the developmental history of neurons which govern their spatial 
organization as well as their inter-connectivity. The latter, as
we have shown above, is guided by several types of homophily, i.e., the 
tendency of neurons which are similar
in terms of certain features - viz., process length, lineage, birth-time
and bilateral symmetry - to be connected via synapses or gap junctions.
We shall now see how the functional identities of neurons are related to
their developmental histories. In particular, we show that
classes of neurons distinguished by their (i) functional identity (viz., 
sensory, motor and interneurons), (ii) functional role in the mesoscopic structural
organization of the network and (iii) membership in distinct 
functional circuits, strongly influences the temporal order of
their appearance in the developmental chronology of the nervous system.

\begin{figure*}[t]
\centerline{\includegraphics[width=\linewidth]{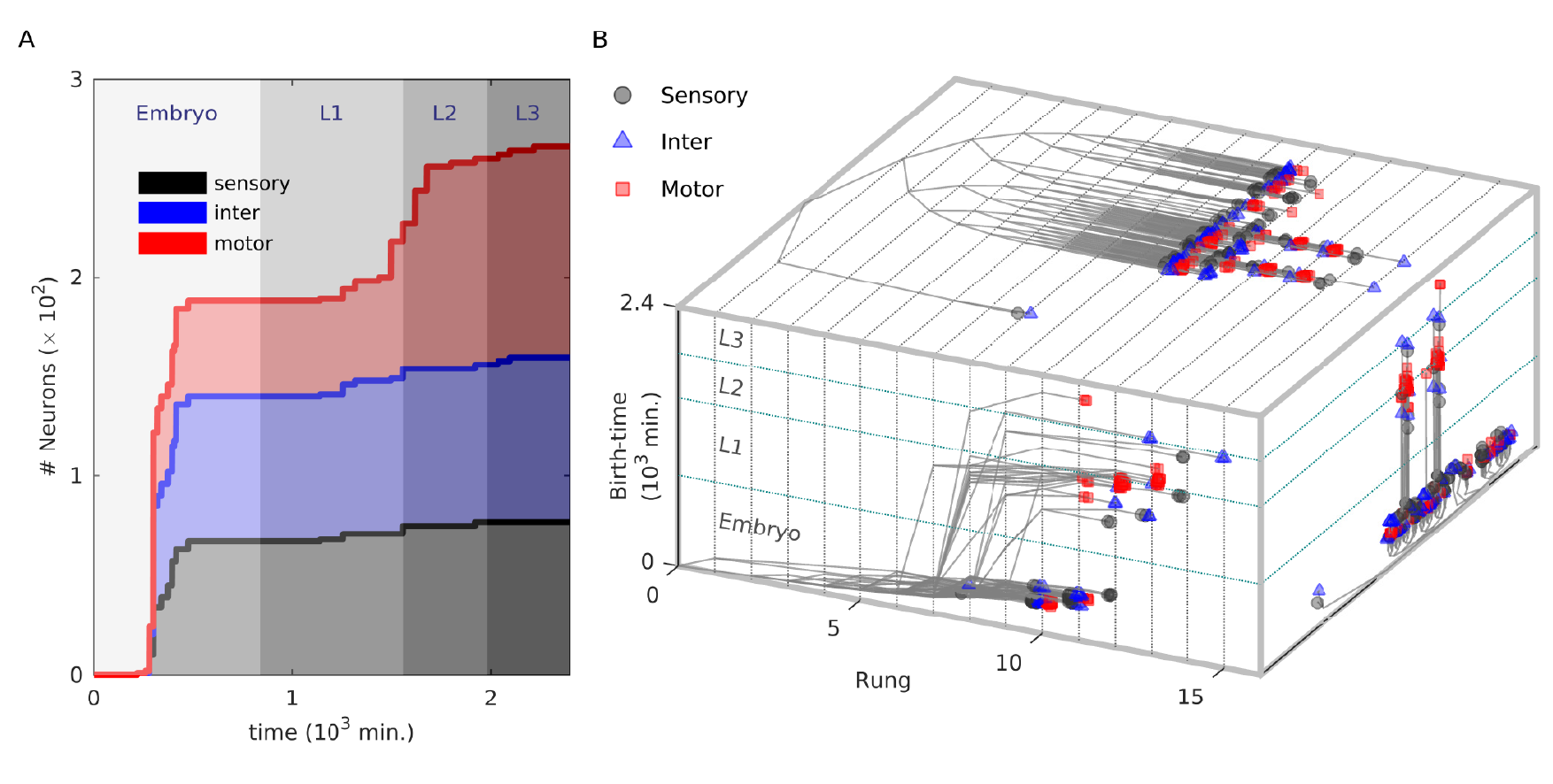}}
\caption{\textbf{Developmental histories of neurons show a bifurcation into early and late branches, with a predominance
of motor neurons in the latter.}  
(A) Bulk of the sensory and interneurons appear early, i.e., during the embryonic stage, while a large fraction of
motor neurons differentiate much later (L2 or L3) during development. 
(B) Planar projections of a three-dimensional representation of the 
developmental history of the entire somatic nervous system
of {\em C. elegans}. Different colors and symbols have been used to denote distinct neuron types (viz., sensory,
motor and interneurons). 
The projection on the top surface
shows the lineage tree with
branching lines connecting the single cell zygote (shown at rung 0) to each of the differentiated neurons
located on their corresponding rungs. 
At higher rungs ($>11$) we see that the differentiated cells are tightly clustered into two
bundles of branches with a predominance of motor neurons (also seen in the chrono-dendrogram projection shown
at the right face of the base).
We note the absence of segregated clusters comprising exclusively the same 
functional type of neurons (viz., sensory, motor or inter), 
suggesting that the
progenitor cell can give rise to neurons of different types.
This in turn implies that commitment to a particular neuron function 
occurs quite late in the sequence of cell divisions.
The projection along the base (left face) shows trajectories representing the developmental history of
each final differentiated neuron, indicating the time of each cell division starting from the zygote along with
the corresponding rung.
For the first few rungs, cell division across different lineages appear to be synchronized and
occur at regular time intervals, which is manifested as an almost linear relation between
time of division and rung. However, between 
rungs 6-9, we observe a bifurcation of the trajectories into two clusters widely separated in time. 
One of these comprises cells 
which differentiate in the embryonic stage 
(termed as the ``early branch'') while the other consists of cells that differentiate much
later (``late branch''). This is manifested in a bimodal distribution of birth times
for neurons occurring in rungs $\geq 10$.
In contrast to the regularly spaced cell divisions in the early branch,  
the trajectories belonging to the late branch are widely dispersed, with
relatively little correlation
between birth time of neurons and their rungs. 
}
\label{fig6}
\end{figure*}

{\em Functional types.}
One of the simplest classifications of neurons is according to their position
in the hierarchy along which signals travel in the nervous system. Thus,
{\em sensory} neurons receive information from receptors located
on the body surface of the organism and transmit them onward to 
{\em interneurons}, which allow signals arriving from different parts
to be integrated, with appropriate response being eventually communicated to
{\em motor} neurons that activate effectors such as muscle cells.
In the mature {\em C. elegans} somatic nervous system, the motor
neurons form the majority ($106$), while sensory ($77$) and 
interneurons ($83$) are comparable in number. The remaining neurons
are polymodal and cannot be uniquely assigned to a specific functional type.
In Fig.~\ref{fig6}~(A) we show how the sub-populations corresponding to
each of the distinct functional types evolve over the course of 
development of the organism.
We immediately note that while the bulk of the sensory and interneurons
differentiate early, i.e., in the embryonic stage, followed by a 
more gradual appearance of the few remaining ones in the larval stages, 
more than half of the motor
neurons appear much later after hatching. 
Moreover, of the $48$ motor neurons which appear early, 
approximately half ($23$) 
belong to the nerve ring while the rest are in the ventral cord, where they
almost exclusively innervate dorsal muscles
(the positions of neurons, classified according to
function type and birth time, is shown in 
Supplementary Material, Fig.~\ref{S11}). 
On the other hand, the $58$ late-born motor neurons primarily belong to
the ventral cord (with only $4$ appearing in the nerve ring). In addition, the 
majority of them ($41$) innervate ventral body muscles
(see Supplementary Table~\ref{T3} for details). The few ($11$) late-born
motor neurons that do innervate dorsal muscles differ from the early-born
ones in that they do not have complementary partners and bring about
asymmetric muscle activation~\cite{tolstenkov2018}.
This early innervation of dorsal muscle but late, larval-stage
innervation of ventral muscles could embody developmental
constraints that deserve further exploration in the future.

Having looked at how neurons emerge according to their functional
type at different times and at different locations
in the physical space described by the body of the worm, we
now consider the appearance of such neurons in the developmental
space defined by lineage and birth time [Fig.~\ref{fig6}~(B)].
The projections of the chrono-dendrogram
that are shown on the top and the extreme right surfaces, both
correspond to representations of the lineage tree that are demarcated by
rung and birth time, respectively. We note immediately that the
developmental trajectories of the neurons appearing in the late burst 
of development are clustered into two distinct branches
that originate in an early
division across left/right axis of the common ABp blastomere
(i.e., cells in one branch originate from ABpl, while those
in the other emanate from ABpr).
Unlike the case seen for neurons belonging to a specific ganglion, we
observe that neurons of the same functional type do not form
localized clusters in the tree that would have suggested a common
ancestry.
Thus, progenitor cells can give rise to neurons of each of the different 
functional types, suggesting that the
commitment to a sensory/motor/interneuron fate happens 
later in the sequence of divisions during development. 

The projection on the remaining bounding surface (left face of the base) 
shows the trajectories 
followed by cells to their eventual neuronal fate across
a space defined by the rung of the lineage tree along one axis and 
the time of cell division along the other.
These trace the developmental history of the entire ensemble of neurons 
comprising the somatic nervous system. We observe that in the early phase
of embryonic stage (corresponding to rungs $\leq 6$) 
there is a linear relation between the time at which
a cell divides and the rung occupied by the 
resulting daughter cells.
This implies that cell divisions across different branches of the lineage
tree occur at regular time intervals in a synchronized manner.
Following this, we observe that the 
trajectories bifurcate and cluster into two branches that are widely 
separated in time. The `early branch', which results in cells
differentiating to a neuronal fate much before hatching, continues
to follow the trend seen in the earlier rungs. However, several
progenitor cells (that can occur in rungs ranging between 6 and 9) 
suspend their division for extremely long times, i.e., until after 
hatching. These comprise the `late branch' where the final neuronal cell
fate is achieved in the larval stages (L1-L3). The occurrence
of these two branches gives rise to the 
bimodal distribution of birth-times shown in Fig.~\ref{fig2}~(B).
In contrast to the regular, synchronized cell divisions across different
lineages seen in the `early branch', the `late branch' exhibits a relative
lack of correlation between rung and birth time, manifested as a wide
dispersion of trajectories followed by individual cells.
We note that the majority of differentiated neurons that eventually result
from the late branch are motor neurons, which corresponds to the
late increase in the subpopulation of motor neurons seen 
in Fig.~\ref{fig6}~(A).
Although there is little information as to when synapses form,
the late appearance of the majority of the motor neurons could suggest
that stimuli from neighboring neurons are playing an important role
in shaping their connectivity in comparison to that of
sensory and interneurons that are primarily guided by
molecular cues.

{\em Mesoscopic functional roles.}
Turning from the intrinsic features of neurons to the properties 
they acquire as a consequence of the network connection topology,
we observe that it has been already
noted that neurons that have a large number of connections are born 
early~\cite{kaiser2011}. This could possibly arise as a result of
the longer time available prior to maturation of the organism 
for connections between these early born neurons to be formed 
with other neurons, including those that differentiate much later.
However, as many neurons which have relatively fewer connections are
also born in the early stage, there does not seem to be a simple relation
between the degree of a neuron and its place in the developmental chronology.
To explore in more depth how the connectivity of a neuron is related to the
temporal order of their appearance,
we therefore consider the role played by it
in the mesoscopic structural organization 
of the network.

\begin{figure*}[t]
\centerline{\includegraphics[width=\linewidth]{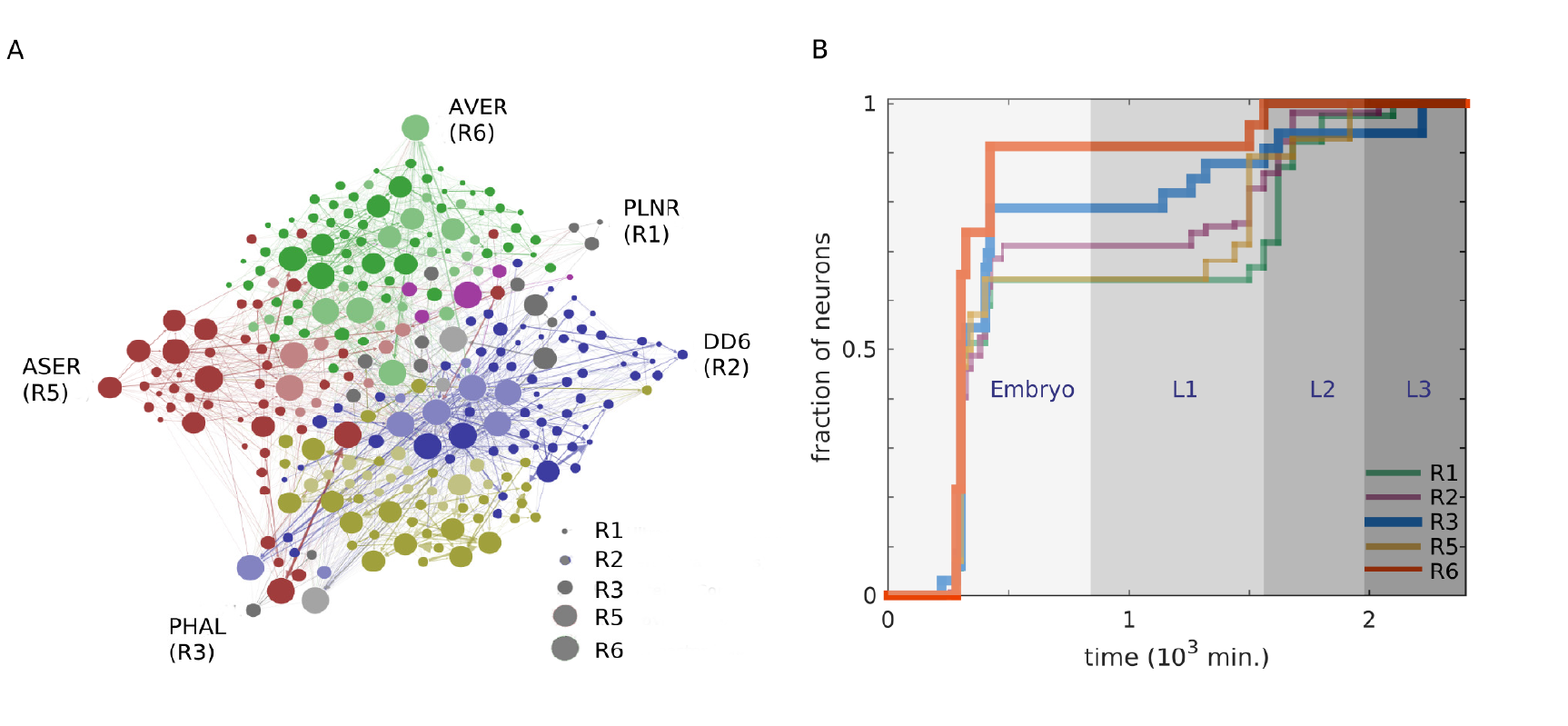}}
\caption{\textbf{Neurons functioning as connectors between different
network modules lead in development.} 
(A) Schematic representation of the network of neurons belonging to 
the somatic nervous system of {\em Caenorhabditis elegans}, 
indicating the role of each neuron (indicated by the node size, see legend) in the mesoscopic structural 
organization
of the network. 
This organization is manifest in the partitioning of the entire network
into six structural modules~\cite{rkpan2010} which are 
characterized 
by relatively dense connections among neurons in each
module compared to the connections between neurons belonging to different modules (node color representing 
the identity of a module to which a neuron belongs).
Within each module, neurons can be further distinguished into
those which have significantly higher number of connections to neurons within
their own module (hubs) and those which do not (non-hubs).
According to their intra- and inter-modular connectivity,
every neuron is then classified into one of seven possible categories (see
Methods), viz.
R1: ultra-peripheral (non-hub nodes with all 
their connections confined to their own module), 
R2: peripheral (non-hub nodes 
with most of their connections occurring within their module), 
R3: satellite connectors (non-hub nodes having
with many connections to other modules), R4: kin-less (non-hub nodes 
with connections distributed uniformly among all modules), R5: provincial hubs 
(hub nodes with a large majority of connections within their module), 
R6: connector hubs (hub nodes with 
many connections to other modules) and R7: global hubs (hub nodes with 
connections distributed uniformly
among all modules). 
One representative neuron from each of the categories is separately 
indicated with a label identifying them by name (note that there are 
no neurons in the {\em C. elegans} somatic nervous system
which belong to categories R4 or R7). 
Neurons which function as connectors, e.g., 
AVER (R6) and PHAL (R3), are seen to have links
to neurons belonging to many different modules (as indicated by the node
color of their network neighborings)
while neurons belonging to other categories
are connected
predominantly to neurons within their own modules (indicated by
their network neighborhood being almost homogeneous in terms
of node color). Neighbors of labeled neurons are either shown
clustered around them (for PLNR, DD6, PHAL and ASER) or 
indicated by a lighter shade of node color (for AVER). 
(B) Distribution of differentiation times of neurons belonging to 
the different network functional role categories 
indicate that the development of those functioning as 
connectors (viz., R3 and R6)
lead the other classes of neurons in the embryonic, as well as, L1 stages.
In particular, more than $90 \%$ of connector hubs have appeared before 
hatching, while for the non-connector categories (R1, R2 and R5),
$70\%$ or less of their members would have differentiated by that time.
}
\label{fig7}
\end{figure*}

Specifically, we focus on the six previously 
identified {\em topological modules}
of the {\em C. elegans} neuronal network, which are groups of neurons
that have markedly more connections with each other than to neurons
belonging to other modules~\cite{rkpan2010}.
We classify all the neurons by identifying their
function in terms of linking the elements belonging to a module, as well as,
connecting different modules to each other~\cite{Guimera2005}. This is done by measuring 
(i) how significantly well connected a neuron is to other cells in its
own module by using the within-module degree $z$-score, and (ii) how 
dispersed the connections of a neuron are among the
different modules by using the 
participation coefficient $P$~\cite{Guimera2005a}.
Cells are classified as hub or non-hub
based on the value of $z$ (see Methods for details).
The hubs can be further classified based on the value of $P$
as (R5) {\em local or provincial} hubs, that
have most of their links confined
within their own module and (R6) {\em connector} hubs, that have a
substantial number of their connections distributed among other
modules. 
The measured value of $P$ is also used to divide the non-hub neurons into
(R1) {\em ultra-peripheral} nodes, which connect only to
members of their own module,
(R2) {\em peripheral} nodes, most of whose links are restricted within
their module and (R3) {\em satellite connectors}, that link to a reasonably
high number of neurons outside their module. 
Fig.~\ref{fig7}~(A) shows the roles (indicated by node size) 
played by each neuron in the somatic
nervous system of {\em C. elegans} using a schematic representation of the
network.
In principle, while it is also possible to have (R7) {\em global} hubs
and (R4) {\em kinless} nodes, viz., hub and non-hub nodes
that may connect to other neurons homogeneously, regardless of their module, 
none of the neurons appear to play such roles in the network.

Earlier investigation~\cite{rkpan2010} has already established that
the connector hubs are crucial in coordinating most of the vital
functions that the {\em C. elegans} nervous system has to perform.
Indeed, $20$ of the $23$ neurons having this role are seen to occur in
one or more functional circuits (discussed later).
Their importance to the network is further reinforced by observing
from Fig.~\ref{fig7}~(B) that all but two of the neurons belonging
to the R6 category appear early in the embryonic stage, and even the
remaining ones, viz., AVFL/R (discussed earlier in the context of
bilaterally symmetric neurons), differentiate by the end of L1 stage. 
By contrast, all other functional role categories have a much smaller 
fraction of their members appear in the early burst of development
and have to wait till the L2 or L3 stage for the development
of their full complement. In particular, we notice that the provincial
hubs, despite having a relatively high degree, lag behind
not only the connector hubs but also the satellite connectors
(that have much lower degree) for most of the developmental period.
This suggests that more than the degree, it is the distribution of
the connections of a neuron among the different modules (quantified
by the participation coefficient $P$), and thus
its functional role in coordinating activity across different parts
of the network, which is an important determinative factor for its 
appearance early in the
developmental chronology of the nervous system.

\begin{figure*}[t]
\centerline{\includegraphics[width=0.9\linewidth]{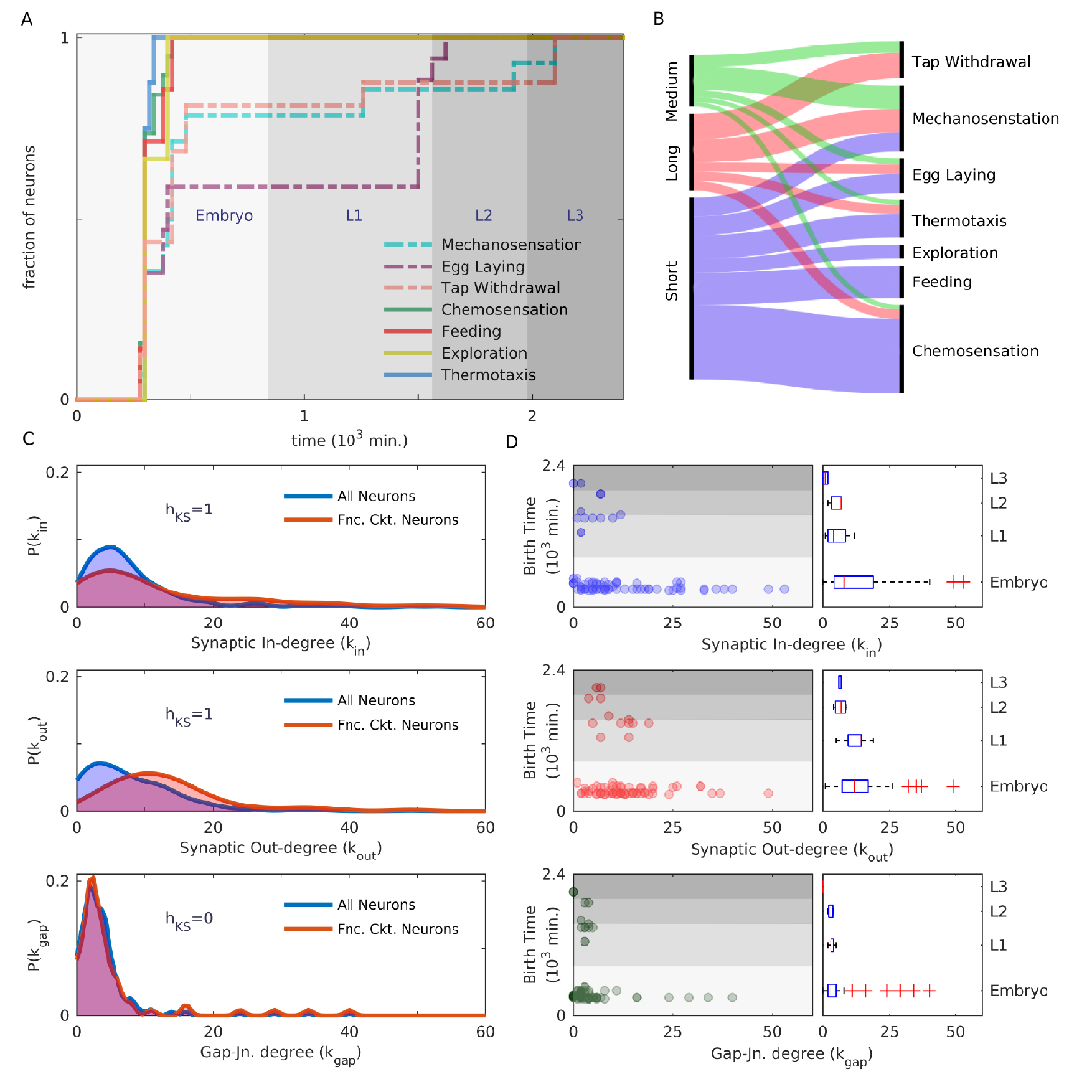}}
\caption{\textbf{The developmental duration of functional circuit neurons 
are strongly indicative of their process length and connectivity.}
(A) Distribution of differentiation times of neurons that belong to
any of seven previously identified functional circuits:
(i) mechanosensation, (ii) egg laying, (iii) tap withdrawal,
(iv) chemosensation, (v) feeding, (vi) exploration and (vii) thermotaxis.
Note that the entire complement of neurons belonging to the functional
circuits corresponding to (iv)-(vii) [shown using solid lines] 
have differentiated early in the embryo stage, 
while those for (i)-(iii) [shown using broken lines] are completed much 
later, viz., in the L2 and L3 stages.
(B) The distribution of neurons having 
short, medium and long processes
[indicated at left],
among the different functional circuits [right].
We note a correlation between the morphological feature of neurite 
length and the development time of functional circuit neurons, viz., 
those in early developing circuits (iv-vii) predominantly have short processes,
while those in later developing circuits (i-iii) mostly have medium
to long processes.
(C) Comparison between the distributions of the number of incoming and
outgoing synaptic connections ($k_{in}$ [top panel] and $k_{out}$ [middle
panel], respectively), as well as, gap junctions ($k_{gap}$ [bottom
panel]) of neurons in the entire somatic nervous system (blue) and 
of the subset of functional circuit neurons (red).
We note that the distribution of synaptic connections (both incoming
and outgoing) for the functional circuit neurons 
is significantly different from that
for the entire network, as indicated by the result of a two-sample 
Kolmogorov-Smirnov test at $5\%$ level of significance ($h_{KS}=1$),
but not for gap junctions ($h_{KS}=0$).
(D) Dispersion of $k_{in}$ (top panel), $k_{out}$ (middle) and $k_{gap}$
(bottom) for the functional
circuit neurons differentiating
at various times is shown in terms of the adjoining box plots where neurons
are clustered into four groups according to the developmental stage during 
which they are born, viz., Embryo, L1, L2 or L3.
In general, the distributions are far more broad for the early born neurons 
(Embryo) compared to those born later (L1-L3).
Focusing on the functional circuit neurons that
develop in the embryonic stage, we note that the distribution of
incoming connections is more skewed than that for outgoing connections.
The distribution of gap junctions is even more skewed, with outliers
lying very far from the median.}
\label{fig8}
\end{figure*}
{\em Membership in functional circuits.}
In order to delve deeper into a possible association between the function(s)
that a neuron performs in the mature nervous system and its developmental 
characteristics, specifically its place in the temporal sequence
of appearance of the neurons, we now focus on several previously 
identified functional circuits of {\em C. elegans}. These are groups of
neurons which have been identified by behavioral assay of
individuals in which the cells have been removed (e.g., by laser ablation).
As their absence results in abnormal or impaired performance of specific
functions, these neurons are believed to be crucial for executing those
functions, viz., 
(F1) mechanosensation~\cite{sulston85, wicks95, sawin96}, 
(F2) egg laying~\cite{waggoner98, bany2003}, 
(F3) thermotaxis~\cite{mori95},
(F4) chemosensation~\cite{troemel97},
(F5) feeding~\cite{White1986, sulston85, gray2005},
(F6) exploration~\cite{White1986, sulston85, gray2005}
and (F7) tap withdrawal~\cite{wicks95, wicks96}.
Note that several neurons belong to multiple functional circuits.
Fig.~\ref{fig8}~(A) shows that one can classify these seven functional circuits into two
groups based on whether or not all the constituent neurons of a circuit appear during the early 
burst of development in the embryonic stage. Thus, while circuits for F3-F6 (shown using solid lines
in the figure) have their entire complement of cells differentiate prior to hatching,
circuits for F1, F2 and F7 lag behind (broken lines), with less than $60\%$ of the egg laying 
circuit having appeared 
at the time of hatching. Indeed, for the entire set of neurons for the latter circuits to emerge
one has to wait until 
the much later L2 (for F2) or L3 (for F1 and F7) stages [note that out of the 16 neurons in the 
F7 circuit, 15 are common to those belonging in the F1 circuit, making the former almost a subset of the
latter].
The temporal order in which the circuits appear makes intuitive sense in that, the functions
that are vital for survival of the organism at the earliest stages (such as thermotaxis
or chemosensation) have all the components of their corresponding circuits completed 
much earlier than those functions such as egg laying which are required only in
the adult worm. 

An intriguing relation between process lengths of neurons and their occurrence in 
different functional circuits is suggested by Fig.~\ref{fig8}~(B), from which we see that
circuits which have their entire complement of neurons differentiate early, viz., F3-F6, are
dominated by neurons having short processes.  In contrast, circuits such as F1, F2 and F7 
that take much longer to have all their members appear comprise a large number of neurons
with medium or long processes. This association between a morphological feature 
(viz., neurite length) of a functionally important neuron
and its time of appearance suggests a possible connection with the process
length homophily, viz., preferential connection between neurons having short
processes, mentioned earlier. Neurons with short processes that belong to 
the ``early'' functional circuits are
mostly chemosensory or interneurons that are all located in the head region.
To perform their task these neurons only need to connect to each other, whose
cell bodies are mostly in close physical proximity of each other. 
Moreover, having their synapses localized within a small region allows 
them to be activated by neuromodulation through diffusion of peptides and other 
molecules~\cite{bargmann2012,Bentley2016}. This assumes significance in light of our observation
that process length homophily between short process neurons is marginally enhanced
in the head. The value of $Q$, a quantitative measure of
homophily introduced earlier, is 0.11 (for synapse, for gap junctions it is $0.12$) 
for early born short process neurons which have their
cell bodies located in the head region. In contrast, when we consider
all short process neurons, $Q$ is $0.067$ for synapse and $0.073$ for
gap junction, respectively.
Thus, the process length homophily we reported earlier could arise in short process
neurons because of functional reasons.

We shall now see how consideration of functional circuits help in obtaining a deeper understanding 
of the nuanced relation between the degree of a neuron and the time of its birth
that was discussed above (in the context of mesoscopic functional roles of neurons). 
As seen in Fig.~\ref{fig8}~(C), neurons belonging 
to the functional circuits show a significantly different distribution for the number of
synaptic connections (both incoming and outgoing) from that of the entire system, as indicated by 
the results of two-sample Kolmogorov-Smirnov test (test statistic $h_{KS}=1$) 
at $5\%$ level of significance.
Thus, the set of functionally critical neurons - which, on average, have a larger number
of connections than a typical neuron in the somatic nervous system - may need to be
treated separately from the other neurons when we examine how synaptic degree correlates with 
birth time.
In contrast, their gap junctional degree distributions cannot be distinguished from
that of all neurons, as indicated by the result ($h_{KS}=0$) for the statistical test
of significance.

Considering only the neurons that appear in functional circuits, we observe that
most of the neurons having a large number of synaptic connections (particularly, incoming
ones) do tend to appear early [Fig.~\ref{fig8}~(D)]. 
On comparing the distributions of synaptic in-degree separately for early and late appearing
functionally critical neurons (see Supplementary Material, Fig.~\ref{S12})
we note that their difference is indeed statistically significant. However, a deeper 
scrutiny shows that the significant deviation between the two is a result of the
occurrence of the $14$ largest in-degree neurons in the group that is born early, all of which turn
out to be connector hubs (described earlier). On removing these, the in-degree distributions for
early and late born neurons belonging to the functional circuits become indistinguishable.
Thus, the distinction between the two sets of neurons in terms
of their degree can be traced to the distinct functional roles, rather than simply
their order of appearance in the developmental chronology. Moreover, when we consider
the rest of the neurons of the somatic nervous system, the early and late born ones 
cannot be distinguished in terms of their in-degree distributions. Thus, birth time
does not appear to be a significant determinant for the connections that are received by a 
post-synaptic neuron.

When we consider the distribution of the synaptic out-degree we see a very
different result. The distributions for the early and late born functionally
critical neurons turn out to be statistically indistinguishable (despite the
appearance of a few extreme outliers such as the command interneurons AVAL/R and 
PVCL/R). In contrast, the rest of the neurons show a much broader distribution (statistically
distinguishable using a two-sample Kolmogorov-Smirnov test) 
for the neurons that are born early, compared to 
those which are born late. This is consistent with the assumption that pre-synaptic neurons 
that exist for a longer period during development, are able to form many more connections
than those neurons which appear later (the latter presumably having less
time to form connections before the maturation of the nervous system). From this
perspective, it is thus striking that the late born functionally critical neurons
have as many connections as they do (making them statistically
indistinguishable from the early born set), and is possibly related to their
inclusion in the functional circuits.

When we consider the gap-junctional degree distributions, we observe that there is no
statistically significant difference between the distributions for early
and late born neurons, whether they be functionally critical or other neurons.
The box plots showing the nature of the distribution at different developmental
stages are all fairly narrow [bottom panel of Fig.~\ref{fig8}~(D)], even though 
the embryonic one shows several outliers with the four farthest ones being the
command interneurons AVAL/R and AVBL/R that appear in four functional circuits, viz., those
of mechanosensation, tap withdrawal, chemosensation and thermotaxis. 
These, in fact, correspond to the outlying peaks 
of the $k_{gap}$ distribution, located on the extended tail at the
right of the bulk [bottom panel of Fig.~\ref{fig8}~(C)].
Indeed, the outliers in each of the distributions (for $k_{in}$, $k_{out}$ and $k_{gap}$),
that appear only at the embryonic stage, almost always happen to be
command interneurons. Of these, AVAL/R are common across the distributions and the fact that they 
occur in four of the known functional circuits underlines the relation between 
function, connectivity and the temporal order of appearance of neurons that we have
sought to establish in this paper.

\section{Discussion}
The nervous system, characterized by highly organized patterns of
interactions between neurons and associated cells, is possibly the 
most complex of all organ systems that is assembled in an animal
embryo over the course of development~\cite{Wolpert2011}.
For this neural network to be functional, it is vital that the
cells are able to form precisely delineated connections with 
other cells that will give rise to specific actions. This raises
the question of how the ``brain wiring problem'' is resolved
during the development process of an organism. In addition
to the processes of cellular differentiation, morphogenesis and
migration that are also seen in other tissue, cells in the nervous 
system are also capable of activity which modulates the development
of the neighboring cells they may interact with. Processes extending
from the neuronal cell bodies are guided towards designated targets
by molecular cues, and the resulting connections are subsequently
refined (e.g., by pruning) through the activity of the cells themselves.
In this paper we have looked at 
a more abstract ``algorithmic'' level of guiding principles that can
help in connecting the details of cellular wiring at the 
implementation level of molecular mechanisms with the final result,
viz., the spatial organization and connection topology of an 
entire nervous system. Using the relatively simple nervous system 
of the model organism {\em Caenorhabditis
elegans}, whose entire developmental lineage and connectivity are 
completely mapped, we have strived to show how development itself 
provides constraints for the design of the nervous system. 

One of our key findings is that neurons with similar attributes,
specifically, (i) the lengths of the processes extending from the cell body
(short/ medium/ long),
(ii) the birth cohort to which they belong (early/ late), (iii) the extent
of shared lineage and (iv) bilateral symmetry pairing (left/ right), 
exhibit a significant preference for connecting to each other (homophily).
Moreover, excepting for homophily on the basis of lineage relations (which 
is seen for synaptic connections only), all other types of homophily are 
manifested by both the connection topology of the network
of chemical synapses, as well as, that of electrical gap-junctions,
despite the fundamental differences in the nature of these distinct
types of links.

We have already discussed earlier a 
plausible mechanism by which homophily based on lineage 
would be observed only in the case of synapses. This is based
on the hypothesis that
synaptogenesis occurs much earlier than the
formation of gap junctions during development. As neurons
are displaced from their initial locations while retaining the
synaptic connections that have formed already, cells that share
common lineage tend to move apart. Thus, when gap-junctions form
much later between adjacent cells, the connected neurons may
have quite different lineages - disrupting any relation between
lineage distance and probability of connection via gap-junction.
An alternative possibility 
that may also explain the specificity of lineage homophily to synapses 
is related to the suggestion that synaptogenesis could be guided by 
cellular labels that are specified by a combinatorial code
of neural cell adhesion proteins~\cite{Emmons2016}. 
In this scenario, cells that are close in terms of their lineage
will be likely to share several of the recognition molecules that will
together determine the label code. Thus, if a sufficiently large
number of these determinants match each other, it could promote
synapse formation between such cells, resulting in
lineage homophily.
We would also like to note that, apart from playing an important role in
determining the topological
structure of the synaptic network, lineage relations between neurons
also appear to shape the spatial organization of neurons by segregating
them into different ganglia.

In addition to investigating the probability that a connection will occur
between a pair of neurons during development, our study also considers how 
the distance between cell bodies of the neurons thus connected is distributed.
Our results suggest that for synapses, the process length of the
pre-synaptic neuron is a decisive factor in determining the
separation that is allowed between the neuronal partners.
Birth time also appears to play a role, particularly, in the case of
long-range connections, i.e., between neurons whose cell bodies are
separated by more than two-thirds of the worm body length. Specifically,
such connections occur between pre-synaptic neurons that are born early
and post-synaptic neurons that are born late, much more often than is expected
by chance.
This suggests the existence of an active process
for the formation of such long-range connections,
for example, using fasciculation as an axon guidance 
mechanism~\cite{White1986,Kaiser2017}.
The latter involves a few pioneer neurons
with long process lengths acting as supporting pathways
that guide axons of the later developing neurons. 
This may also underlie a triadic closure-like phenomenon in the network~\cite{newmanbook,kleinbergbook} 
(viz., two neurons having links to one or more common neighbors that have an increased likelihood
of being connected to each other). Such a process is known to yield strongly clustered networks
with high communication efficiency~\cite{davidsen2002,brot2015} and could be 
responsible for the appearance of the so-called ``common neighbor rule'' 
that has been reported in the {\em C. elegans} connectome~\cite{azulay2016}.

Our results also indicate that the temporal sequence 
in which neurons appear during development of the nervous system
is linked to their functional identities. The simplest of these
identities is simply the basic functional type of a neuron, viz., 
whether it is sensory, motor or interneuron.
The neurons belonging to these different types not only differ
in terms of the cells they connect to (for instance, only motor neurons 
connect to muscles and sensory neurons are the only ones to receive
connections from receptors, while interneurons connect to all types of
neurons) but also in their molecular inventory. 
While their lineage does not show any significant differences, the different
functional types of neurons do appear to segregate to a certain extent
in terms of their time of appearance. Specifically, we find that the bulk
of neurons that are born in the late, post-embryonic burst of
development are motor neurons. Our results suggest that there 
may be temporal cues that appear late in the process of development
which are responsible for the specialization
of neurons into different functional types according to their time of
birth.

At a higher, mesoscopic level of organization of the network structure,
we have considered the functional role of neurons in coordinating
the activity of different topological modules of the network. We show that
this allows us to obtain a much more nuanced picture of how the number
of connections that a neuron has with other neurons, affects its place 
in the temporal
sequence in which neurons appear during development. Thus, rather
than a simple case of just the degree (the total number of connections) of a
neuron deciding its precedence in the sequence, it is 
actually the connector neurons (connector hubs, as well as,
satellite connectors) which are the earliest to appear.
We also examine in detail the subset of neurons that have been identified 
as belonging to one or more functional circuits in the {\em C. elegans}.
We observe that membership of a specific functional circuit does determine
the order in which these neurons appear, with certain circuits,
such as those responsible for chemosensation, emerging early (before
hatching) while others circuits, such as those for mechanosensation and 
egg-laying, appear much later (after hatching). In turn, the time of
appearance of functional circuits determines to an extent the morphological 
properties, such as the process lengths, of their constituent neurons. 
   
The observations reported in this paper are an attempt at resolving the 
``wiring problem'' for the {\em C. elegans} nervous system
by focusing at a level that is intermediate between the molecular 
mechanism-level details of developmental processes and the 
resulting structural organization of the entire somatic nervous system. 
Specifically, we have attempted to uncover general
``algorithmic'' principles governing 
the design of the neuronal network, which will allow linking
the complicated molecular machinery involved to the equally complicated
spatial and topological description of the nervous system. The next step
in this approach will involve delineating exactly how these governing principles
(such as the various types of homophily) are implemented by
molecular mechanisms, and how genetics may be relating 
the temporal sequence of appearance
of neurons to their functional identities.
Experimental and theoretical progress towards this direction 
would enable us to achieve a seamless 
understanding of nervous system development involving different scales.

%

\section{Materials and Methods}
\subsection{Data}
\label{mm_data}
{\em Connectivity and lineage.}
We have used information about the network connectivity and
lineage distance between $279$ connected neurons of 
the {\em C. elegans} somatic 
nervous system from the database published in Ref.~\cite{varshney2011},
accessible from an online resource for behavioral and structural anatomy of
the worm~\cite{wormatlas}. This is an updated and revised version of
connectivity data obtained through serial section electron
micrography that first appeared in Ref.~\cite{White1986}.

{\em Time of birth for neurons.}
We have transcribed the time of appearance of each neuron over the
course of development of the organism from lineage charts provided in 
Refs.~\cite{sulston77, sulston83}. This is provided in Table~\ref{T4}
in the Supplementary Material.

{\em Time of cell-division for progenitor cells}
The information about the time of each cell-division, starting
from the zygote, that occurs over the course of development of the 
{\em C. elegans} somatic nervous system, and which has been
used for generating the chrono-dendrograms shown here, are provided
in Refs.~\cite{sulston77, sulston83}, accessible from an online
interactive visualization application~\cite{wormweb}. 

{\em Physical distance.}
We have used information on the positions of the neurons from the
database reported in Ref.~\cite{choe2004}, accessible online from
\url{https://www.dynamic-connectome.org/}. 
The location information provides coordinates of each neuronal cell body
projected on a two-dimensional plane defined by the anterior-posterior axis
and the dorsal-ventral axis.
 
{\em Neuronal process length.}
Lengths of the processes extending from each cell body has been estimated
from the diagrams of the neurons provided in Appendix 2, Part A of 
Ref.~\cite{wood88} and from an online resource for the
anatomy of the worm~\cite{wormatlas}. 

{\em Ganglia and functional types.}
The information about the ganglion to which a neuron belongs and 
the functional type of each neuron (viz.,
sensory, motor or interneuron) has been obtained from the database
provided in Ref.~\cite{yamamoto91}. 

{\em Functional circuits.}
The identities of the neurons belonging to each of the functional circuits
referred to here is available from the Supporting Information (Table S2) of 
Ref.~\cite{rkpan2010}. 

\subsection{Modularity}
\label{mm_modularity}
A network can be partitioned into several communities or topological
modules, defined such 
that neurons in
a given module have a much higher probability of being connected to
other neurons in the module compared to neurons that do not belong to it, by 
maximizing the modularity value $Q$~\cite{newman2004} associated with 
a given partitioning, viz.,
\begin{equation}
Q = \frac{1}{L}\sum_{i,j}\Big[A_{ij}-\frac{k_i^{in}k_j^{out}}{L}\Big]
\delta_{c_ic_j}.
\end{equation}
Here, $\textbf{A}$ is the adjacency matrix describing the connections
of the network ($A_{ij} = 1$, if neuron $i$ receives a connection from neuron
$j$, and $=0$ otherwise). The in-degree and out-degree of a node $i$ are given
by $k_i^{in}=\sum_{j}A_{ij}$ and $k_j^{out}=\sum_{i}A_{ij}$, respectively.
The total number of links in the network is given by $L = \sum_{i,j}A_{ij}$.
The Kronecker delta function $\delta_{ij}=1$, if $i=j$, and $0$, otherwise.
The indices $c_i, c_j$ refer to the modules to which the neurons $i$
and $j$, respectively, belong. 
For an undirected network, such as that defined by the set of connections
between neurons via gap-junctions, the adjacency matrix is symmetric (i.e.,
$A_{ij} = A_{ji}$) and $k_i^{in}=k_i^{out} = k_i$.
The value of $Q$ expresses the bias that a neuron has to connect to
members of its own community (which could be defined in terms of
any distinguishing characteristic of the cells, e.g., process length),
relative to the null model. The latter corresponds to an 
unbiased, homogeneous network
where the probability of connection between two nodes is proportional
to the product of their respective degrees.

\subsection{Bimodality coefficient}
\label{mm_bimodality}
The bimodal nature of a probability distribution can
be characterized by calculating its bimodality coefficient~\cite{pfister2013}:
\begin{equation}
 BC = \frac{m^2_3+1}{m_4+3\cdot\frac{(n-1)^2}{(n-2)(n-3)}},
\end{equation}
where $m_3$ is the skewness, $m_4$ is the excess kurtosis and $n$ represents
the sample size.  
A distribution is considered to be bimodal if $BC > BC^{*}$ where 
$BC^{*} = 5/9$. This benchmark value corresponds to a uniform distribution,
and if $BC < BC^{*}$, the distribution is considered unimodal.

\subsection{Process length randomization.}
\label{mm_processlengthrand}
To establish statistically significant evidence for process length
homophily, the empirical network is compared with an ensemble
of networks obtained from the empirical one by randomly assigning 
process lengths (short, medium and long) to the neurons while ensuring
that the total number of neurons in each process length category, viz., 
$N_S$, $N_M$ and $N_L$, respectively,
(as well as, all other properties of the network, such as connectivity) 
remains unchanged. In practise, this is done by first partitioning the neurons
into three communities according to process length and ordering the 
neurons in sequence according to the module they belong. Thus, neurons $i=1,
\ldots, N_S$ have short processes, neurons $i=N_S+1, \ldots, N_S+N_M$ have
medium length processes, and neurons $i=N_S+N_M+1, \ldots N_S+N_M+N_L$
have long processes. Then, to create each member of the surrogate ensemble, 
this sequence is randomly permuted and the first
$N_S$ neurons are assigned short process length, the next $N_M$ neurons are 
assigned medium process length and the remaining $N_L$ neurons are assigned
long process length. The modularity $Q$ calculated for networks
with such randomized module membership (corresponding to a null model
where process length homophily is non-existent by design) is 
expected to be small.

\subsection{Network randomization constrained by neuronal process lengths}
\label{mm_netrandom}
An ensemble of surrogate networks is constructed by randomizing the connections
of the empirical network, subject to different constraints.
Each member of the ensemble is constructed by repeatedly selecting
a pair of directed connections, e.g., $p \rightarrow q$ and $u \rightarrow v$,
and rewiring them such that the in-degree and out-degree
of each neuron remains invariant, i.e., $p \rightarrow v$ and $u \rightarrow q$.
If these new connections already exist, this rewiring is disallowed
and a new random selection for a pair of directed connections done.
In addition,
information about the spatial location of 
the cell bodies and that of the process lengths of neurons are used to 
further constrain the connections. This ensures that un-physical connections,
such as between two short process neurons (i.e., each has a process length 
that is less than a third of the body length of the nematode) whose cell
bodies are placed apart by more than 
$2L/3$ ($L$: total body length of the worm),
do not appear through the randomization. 
In practise, this constraint is imposed as follows. In absence of precise
knowledge of the length of each process,
depending on the length process category to which each neuron belongs, 
an uniformly distributed
value (lying between $[0,L/3]$ for short,
between $[L/3,2L/3]$ for medium and $[2L/3,L]$ for long process
neurons) is assigned as the process length for a neuron.
The distance between the cell bodies of a pair of neurons that have
been selected randomly for connection is then
compared against the sum of their process lengths. If the latter is
greater than the former, the connection is allowed, else not.
The rewiring steps are repeated $5 \times 10^5$ times to construct each of
the randomized networks belonging to the surrogate ensemble. The entire
ensemble consists of $100$ realizations of such randomized networks.

\subsection{Lineage randomization}
\label{mm_lineagerandom}
To establish that neurons belonging to the same ganglion are closely
related in terms of their lineage, we compare the properties of the
lineage distance distribution within and between ganglia obtained 
for the empirical network with those obtained upon randomizing the
lineage relations. This is done by repeatedly selecting a pair of neurons
at random
on the lineage tree and exchanging their positions on the tree. This procedure
is carried out $10^4$ times for a single realization. This
ensures that, in the randomized networks, the lineage relation between 
neurons is completely independent of whether they belong to the same ganglion
or not.
In order to compare the
properties of the empirical network with its randomized version,
an ensemble of $10^3$ realizations is considered.
To quantify the deviation of the empirical intra- and inter-ganglionic 
lineage distance distributions from their randomized counterparts,
we measure the $z$-score of the corresponding means and coefficients of 
variation (CV). The $z$-score is a measure for the extent of deviation
of an empirical property $x_{emp}$ from the mean of the randomized
counterparts, $\langle x_{rand} \rangle$, scaled by the standard deviation
of the randomized counterparts, viz.,
\begin{equation}
z=\frac{x_{emp}-\langle x_{rand} \rangle}{\sqrt{\langle x_{rand}^2 \rangle - \langle x_{rand} \rangle^2}}.
\end{equation}

\subsection{Surrogate ensemble for comparison with average cell body 
distance between connected neuronal pairs}
\label{mm_surrogatecellbodydist}
To see whether the distance $d$ between cell bodies of connected pairs of 
neurons [where the members of the pair could belong to either the 
same or different process length categories, viz., short (S), medium (M)
and long (L)]
is distributed in a significantly different manner from that between all pairs
of neurons, we have constructed surrogate ensembles. For each realization
belonging to such an ensemble,
a number of cell body distances is sampled from the set of all distances $D$
between each neuronal pair, such that the sample size is same as the
number of connected neural pairs. The entire ensemble consists of $10^3$
such sampled sets. To see whether the observed difference between
$\langle d_{XY} \rangle$ and $\langle D \rangle_{X,Y}$, where 
$X,Y \in \{ S,M,L \}$, can be explained simply as finite size fluctuation,
we have evaluated the corresponding $z$-scores, viz.,
\begin{equation}
  z_{XY} = \frac{d^{emp}_{XY}-\langle d^{rand}_{XY} \rangle}{\sqrt{\langle (d^{rand}_{XY})^2 \rangle - \langle d^{rand}_{XY} \rangle^2}}. 
\end{equation}


\subsection{Lineage tree rung determination}
\label{mm_rung}
The order of the rung in the lineage tree that a cell belongs to is 
obtained from the lineage information of the cell (available from
Ref.~\cite{wormatlas}). This indicates the series of cell divisions,
starting from AB (which results from the division of the single cell zygote) 
that leads to a particular neuron, e.g., ABprpapaap. The letters a (anterior), 
p (posterior), l (left) and r (right) which follow AB, indicate the 
identity of the progenitor cells that result from subsequent cell
divisions eventually terminating in a differentiated neuron.
As the rung that a neuron belongs to is given by the number of cell divisions 
(starting from the zygote) that leads to the differentiated cell,
we simply count the total number of 
letters (AB is counted as a single letter) specifying the lineage 
of a cell to determines its rung.

\subsection{Stochastic branching model for lineage tree}
\label{mm_lineagetreemodel}
To theoretically describe the generative process leading to the observed lineage
tree for the cells belonging to the {\em C. elegans} somatic nervous 
system, we have used a stochastic asymmetric branching model. 
Starting from the single cell zygote, each cell division leads to
at most two daughter cells, with independent probabilities $P1$ and $P2$ 
($P1 \geq P2$) for the occurrence of each of the two branches.
Thus, based on the probabilities $P1$ and
$P2$, at each step of the generative process any one of the following 
three events can happen: (i) proliferation occurs along both branches, 
(ii) only one branch appears (the other branch leading to either 
apoptosis or a non-neural cell fate), and, (iii) there is no branching so that
a terminal node of the tree is obtained (i.e., the cell differentiates 
into a neuron).
Estimation of $P1$ and $P2$ from the empirical lineage tree suggests 
that proliferation markedly reduces after rung 10. Incorporating
this in the model by decreasing the probabilities $P1,P2$
after rung $10$ results in successive reduction of the branching,
eventually coming to a stop. The ensemble of $10^3$ simulated lineage trees 
produced 
by the process matches fairly well with the empirical lineage tree
in terms of the number of terminal nodes, the distribution of the rungs
occupied by each cell and the distribution of lineage distances between
the differentiated neurons (see Supplementary Material, Fig.~\ref{S4}).

\subsection{Classifying neurons according to their role in the
mesoscopic structural organization of the network}
\label{mm_meso}
The functional importance of a neuron {\em vis-a-vis} its own topological
module (defined above in Sec.~\ref{mm_modularity}), as well as,
the entire nervous system, can be quantified in terms of its
intra- and inter-modular connectivity~\cite{rkpan2010}. 
For this purpose we use the two metrics~\cite{Guimera2005}:
(i) the within module degree $z$-score ($z$) and (ii) the participation 
coefficient ($P$).

In order to identify neurons that have a significantly large number of
connections to the other neurons belonging to their module, we calculate
the within module degree $z$-score defined as
\begin{equation}
z_i = \frac{\kappa^{i}_{c_i}-\langle \kappa^j_{c_i} \rangle_{j\in c_i}}{\sqrt{\langle (\kappa^j_{c_i})^2 \rangle_{j\in c_i}-\langle \kappa^j_{c_i} \rangle^2_{j\in c_i}}},
\end{equation}
where $\kappa^i_c$ is the number of connections that a neuron $i$ has
to other neurons in its community (labeled $c$) and the average 
$\langle \ldots \rangle_{j\in c}$ is taken over all nodes in the community.
Following Ref.~\cite{rkpan2010}, we identify neurons having $z\geq0.7$
as hubs, while the remaining are designated as non-hubs.

The neurons are also distinguished in terms of how many well connected they are to
neurons belonging to other communities. For this purpose we measure the
participation coefficient $P$ of a neuron, which is defined as
\begin{equation}
 P_i = 1 - \sum^m_{c=1}(\frac{\kappa^i_c}{k_i})^2,
\end{equation}
where $\kappa^i_c$, as above, is the number of connections that the neuron has to other neurons in
its own module (labeled $c$) and
$k_i=\sum_c\kappa^i_c$ is the total degree of node $i$.
Neurons that have their connections homogeneously distributed among all modules 
will have a $P$ close to $1$, while $P=0$ if all of their connections are confined
within their module.
Based on the value of $P$, following Ref.~\cite{rkpan2010}
we have classified the non-hub neurons as ultra-peripheral (R1: $P\leq0.05$),
peripheral (R2: $0.05<P\leq0.62$), satellite connectors (R3: $0.62<P\leq0.8$)
and kin-less nodes (R4: $P>0.8$), while hub neurons are segregated into
provincial hubs (R5: $P\leq0.3$), connector hubs (R6: $0.3<P\leq0.75$)
and global hubs (R7: $P>0.75$).


\subsection{Statistics}

{\em Two-sample Kolmogorov-Smirnov (KS) test}~\cite{Massey1951} has been
used to compare between pairs of samples (e.g., the
degrees of neurons belonging to different categories) 
to determine whether both of them
are drawn from the same continuous distribution (null hypothesis) 
or if they belong to different distributions. For this purpose we have 
used the \verb|kstest2| function in {\em MATLAB Release 2010b}, with the
value of the parameter $\alpha$ which determines threshold significance level 
set to $0.05$.

{\em Kernel smoothened density function}~\cite{bowman97} has been used to
estimate the probability distribution functions of different 
quantities (e.g., distances between cell bodies of neurons).
For this purpose we have used
the \verb|ksdensity| function in {\em MATLAB Release 2010b} with a Gaussian
kernel.

\begin{acknowledgments}
We thank Sandhya Koushika for critical feedback, Shakti N. Menon and
Md. Izhar Ashraf for
assistance in collecting data and Upi Bhalla, Aditya
Gilra,
Cathy Rankin, Shawn Xu and
Mei Zhen for helpful discussions.
This work was partially supported by grant BT/PR14055/Med/30/350/2010
of Dept. of Biotechnology, Govt. of India (NC and SS) and IMSc Complex Systems Project (XI Plan) funded
by Dept. of Atomic Energy, Govt. of India (AP).
\end{acknowledgments}

\begin{center}
{\bf \small Author Contributions}
\end{center}
AP, NC and SS conceived the study and designed the methodology, 
AP collected the data, performed the investigation and did the visualization, 
AP, NC and SS analyzed the results and wrote the paper, 
SS supervised the study. 

\begin{center}
{\bf \small Conflict of interest}
\end{center}

The authors declare that they have no conflict of interest.

%

\clearpage
\onecolumngrid
\begin{center}
{\large {\bf SUPPLEMENTARY MATERIAL}}
\end{center}
\setcounter{figure}{0}
\setcounter{equation}{0}
\renewcommand\thefigure{S\arabic{figure}}
\renewcommand\thetable{S\arabic{table}}
\renewcommand{\thesection}{\Roman{section}} 
\renewcommand{\thesubsection}{\thesection.\Roman{subsection}}

\begin{table}[h!]
\caption{{\bf Process length homophily among neurons segregated into groups
comprising cells with long, medium and short processes, respectively.}
The extent of homophily is quantified by the modularity measure $Q$ 
computed over the different classes of neurons (which are considered to
be the communities or modules for the purpose of calculation of $Q$).
The empirical values
are compared with that calculated from the corresponding
surrogate ensemble obtained by randomly shuffling the 
process length categories of the empirical network keeping the network 
connections invariant.
Note that the $Q$ values are significantly higher than that expected by
chance (as seen for
the surrogate ensemble) for the entire network, as well as, individually
for almost all process length categories, suggesting
process length homophily in both synaptic and gap-junction connections
between neurons.\\}
\begin{tabular}{|l|c|c|c|c|}
\hline
        & \multicolumn{2}{c|}{\textbf{Synapse}} & \multicolumn{2}{c|}{\textbf{Gap-junction}} \\ \hline
Process length & Q (empirical)     & Q (randomized)     & Q (empirical)       & Q (randomized)       \\ \hline
Long    & 0.044             & -0.000 $\pm$ 0.006     & 0.051               & -0.001 $\pm$ 0.008       \\ \hline
Medium  & 0.006             & -0.001 $\pm$ 0.004     & 0.01                & -0.001 $\pm$ 0.008       \\ \hline
Short   & 0.067             & -0.001 $\pm$ 0.003     & 0.073               & -0.002 $\pm$ 0.011       \\ \hline
All     & 0.117             & -0.002 $\pm$ 0.010     & 0.134               & -0.004 $\pm$ 0.020       \\ \hline
\end{tabular}
\label{T1}
\end{table}

\begin{table}[h!]
\caption{{\bf For synaptically connected neurons, process length of the 
pre-synaptic neuron primarily decides the average distance between the cell 
bodies.} Statistically significant deviation (measured in terms of $z$-score)
between the average distance
$\langle d \rangle$ of cell bodies in pairs of connected neurons (having
short, medium or long processes) and
the average distance $\langle D \rangle$ between any pair 
of neurons randomly sampled from 
the same process length categories. The latter average is calculated over
a set having the same number of pairs as for the set of 
connected pairs.
Note that except for two cases (pre-synaptic long process to post-synaptic
short process and pre-synaptic medium process to post-synaptic long process, 
shown in bold font), connections
between cells in all other process length categories tend to be much shorter
than that expected by chance, as indicated by $z<0$.\\}
\begin{tabular}{llll}
\hline
\multicolumn{4}{|c|}{\textbf{Synapse}}                                                                                              \\ \hline
\multicolumn{1}{|c|}{}  & \multicolumn{3}{c|}{Pre-synaptic}\\ \hline 
\multicolumn{1}{|c|}{Post-synaptic}       & \multicolumn{1}{c|}{Short}  & \multicolumn{1}{c|}{Medium}       & \multicolumn{1}{c|}{Long}          \\ \hline
\multicolumn{1}{|l|}{Short}  & \multicolumn{1}{c|}{-12.07} & \multicolumn{1}{c|}{-1.79}        & \multicolumn{1}{c|}{\textbf{1.79}} \\ \hline
\multicolumn{1}{|l|}{Medium} & \multicolumn{1}{c|}{-9.13}  & \multicolumn{1}{c|}{-3.52}        & \multicolumn{1}{c|}{-0.29}         \\ \hline
\multicolumn{1}{|l|}{Long}   & \multicolumn{1}{c|}{-7.15}  & \multicolumn{1}{c|}{\textbf{0.9}} & \multicolumn{1}{c|}{-1.19}         \\ \hline
                             &                             &                                   &                                    \\ \hline
\multicolumn{4}{|c|}{\textbf{Gap-junction}}                                                                                              \\ \hline
\multicolumn{1}{|l|}{}       & \multicolumn{1}{c|}{Short}  & \multicolumn{1}{c|}{Medium}       & \multicolumn{1}{c|}{Long}          \\ \hline
\multicolumn{1}{|l|}{Short}  & \multicolumn{1}{c|}{-6.47}  & \multicolumn{1}{c|}{-6.22}        & \multicolumn{1}{c|}{-4.01}         \\ \hline
\multicolumn{1}{|l|}{Medium} & \multicolumn{1}{c|}{-6.22}  & \multicolumn{1}{c|}{-5.16}        & \multicolumn{1}{c|}{-4.47}         \\ \hline
\multicolumn{1}{|l|}{Long}   & \multicolumn{1}{c|}{-4.01}  & \multicolumn{1}{c|}{-4.47}        & \multicolumn{1}{c|}{-2.16}         \\ \hline
\end{tabular}
\label{T2}
\end{table}

\pagebreak

\begin{longtable}[c]{lll}
\caption{{\bf Neurons belonging to the somatic nervous system of 
{\em C. elegans} segregated into those which are born in the early (embryonic) 
and those born in the late
(post-embryonic) developmental
bursts.} The respective lineage information and functional 
description are also provided. Motor neurons are indicated in red. 
We note that motor neurons that appear early mostly innervate dorsal
muscles, whereas, motor neurons that appear late primarily innervate
ventral muscles. All information shown here is 
obtained from WormAtlas~\cite{wormatlas}.}
\label{T3}\\
\hline
\multicolumn{3}{|l|}{{\color[HTML]{343434} \textbf{Early Born Neurons}}}                                                                                                                                                                                                                                \\ \hline
\endfirsthead
\multicolumn{3}{c}%
{{\bfseries Table \thetable\ continued}} \\
\endhead
\multicolumn{1}{|l|}{{\color[HTML]{333333} \textbf{Neurons}}} & \multicolumn{1}{c|}{\textbf{Lineage}}                     & \multicolumn{1}{l|}{\textbf{Description}}                                                                                                                                                           \\ \hline
\multicolumn{1}{|l|}{{\color[HTML]{333333} ADAL}}             & \multicolumn{1}{l|}{AB plapaaaapp}                        & \multicolumn{1}{l|}{Ring interneuron}                                                                                                                                                               \\ \hline
\multicolumn{1}{|l|}{{\color[HTML]{333333} ADAR}}             & \multicolumn{1}{l|}{AB prapaaaapp}                        & \multicolumn{1}{l|}{Ring interneuron}                                                                                                                                                               \\ \hline
\multicolumn{1}{|l|}{{\color[HTML]{333333} ADEL}}             & \multicolumn{1}{l|}{AB plapaaaapa}                        & \multicolumn{1}{l|}{Anterior deirid, sensory neuron}                                                                                                                                                \\ \hline
\multicolumn{1}{|l|}{{\color[HTML]{333333} ADER}}             & \multicolumn{1}{l|}{AB prapaaaapa}                        & \multicolumn{1}{l|}{Anterior deirid, sensory neuron}                                                                                                                                                \\ \hline
\multicolumn{1}{|l|}{{\color[HTML]{333333} ADFL}}             & \multicolumn{1}{l|}{AB alpppppaa}                         & \multicolumn{1}{l|}{Amphid neuron}                                                                                                                                                                  \\
\hline
\end{longtable}

\pagebreak

\begin{longtable}[c]{|l|r|}
\caption{{\bf Time of appearance of neurons in the course of 
development of the {\em C. elegans} nervous system.} The data has been
manually transcribed from lineage charts provided in Refs.~\cite{sulston77, sulston83}.}
\label{T4}\\
\hline
\textbf{Neuron} & \textbf{Birth time (min.)} \\ \hline
\endfirsthead
\multicolumn{2}{c}%
{{\bfseries Table \thetable\ continued}} \\
\endhead
ADAL            & 420                        \\ \hline
ADAR            & 420                        \\ \hline
ADEL            & 420                        \\ \hline
ADER            & 420                        \\ \hline
ADFL            & 280                        \\ \hline
ADFR            & 280                        \\ \hline
ADLL            & 300                        \\ \hline
ADLR            & 300                        \\
\hline
\end{longtable}

\pagebreak
\begin{figure}[h!]
\centerline{\includegraphics[width=0.8\columnwidth]{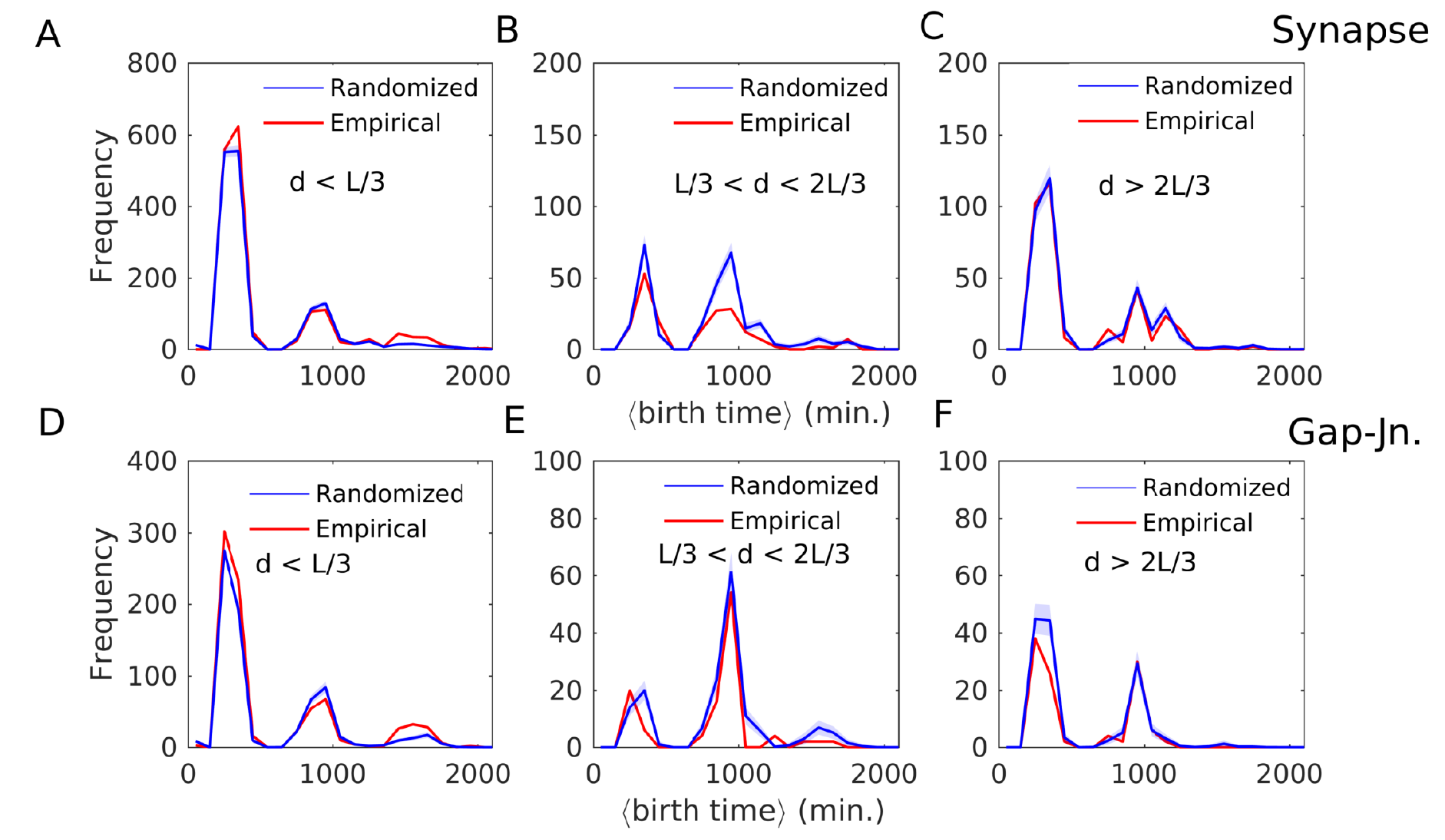}}
\caption{\textbf{Birth cohort homophily is seen specifically for connections
between neurons whose cell bodies are in close physical proximity.} 
Frequency distributions of the mean birth time
for all pairs that are connected via synapses (A-C) or gap-junctions (D-F). 
The distributions for the empirical network (shown in red) are 
compared with distributions obtained from surrogate ensembles of 
randomized networks (blue curve shows the average over $100$ realizations,
the dispersion being indicated by the shaded area).
The latter are constructed from the empirical network by randomly
rewiring the connections while keeping the total number of connections (degree)
for each neuron, the spatial location of its cell body and its process
length unchanged. In addition, to allow only physically possible connections
between neurons, we have imposed process-length constraint which disallow
linking two cells if the distance between their cell bodies is greater than
the sum of their individual process lengths.
The different panels correspond to connections between neurons whose
cell bodies are separated
by distance $d$ which is short ($d<L/3$: A, D), medium ($L/3<d<2L/3$: B, E)
or long ($d>2L/3$: C, F) relative to the total body length of the worm $L$.
As in Fig.~\ref{fig4} in the main text, the trimodal nature of these
distributions arise from three classes of connected neuronal pairs, viz.,
(i) where both cells are born early (i.e., in the embryonic stage), (ii) where 
one is born early while the other late (i.e., in the post-embryonic stage)
and (iii) where both are born late.
Birth cohort homophily is indicated when the 
peaks of the empirical frequency
distribution, corresponding to connections between neurons that are either both
born early or both born late, have significantly higher values than 
the randomized distribution (the latter corresponding to a null model where
connections between cells can occur independent of the time of their birth).
This is seen only in panels (A) and (D), i.e., for connections between
neurons whose cell bodies are located relatively close to each other.
}
\label{figS1}
\end{figure}

\begin{figure}[h!]
\centerline{\includegraphics[width=0.8\columnwidth]{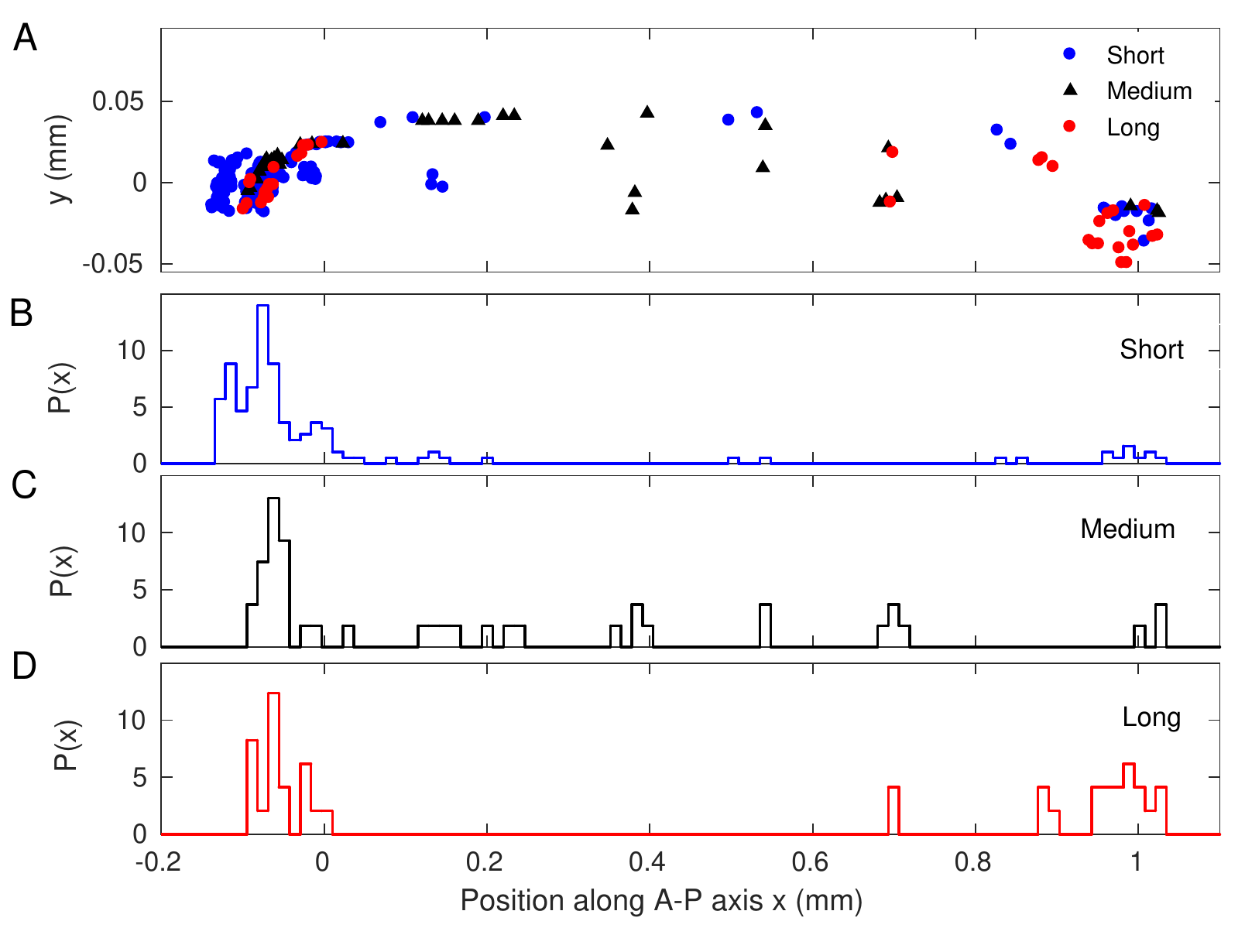}}
\caption{\textbf{Spatial distribution of cell bodies of the neurons
belonging to the somatic nervous system of {\em Caenorhabditis elegans}.} 
(A) Projection of the physical locations of the neuronal cell bodies 
on the two-dimensional plane formed by the anterior-posterior (AP) axis
($x$, along the horizontal)
and the ventral-dorsal axis ($y$, along the vertical).
Cells having short,
medium and long processes are indicated using different symbols.
The animal is oriented such that its head is located near the left end
and tail near the right end of the plane.
(B-D) Probability distributions of the location of the cell bodies along
the AP axis ($x$, measured in mm) for neurons having (B) short, (C) 
medium and (D) long processes. We note that the distributions for
neurons having short and long processes, both have an approximately bimodal
nature. It suggests that most cells of these two types are localized
near either the head or the tail regions, while neurons with medium length
processes are distributed across the body of the worm in a relatively
more homogeneous manner.
}
\label{S2}
\end{figure}

\begin{figure}[h!]
\centerline{\includegraphics[width=0.95\columnwidth]{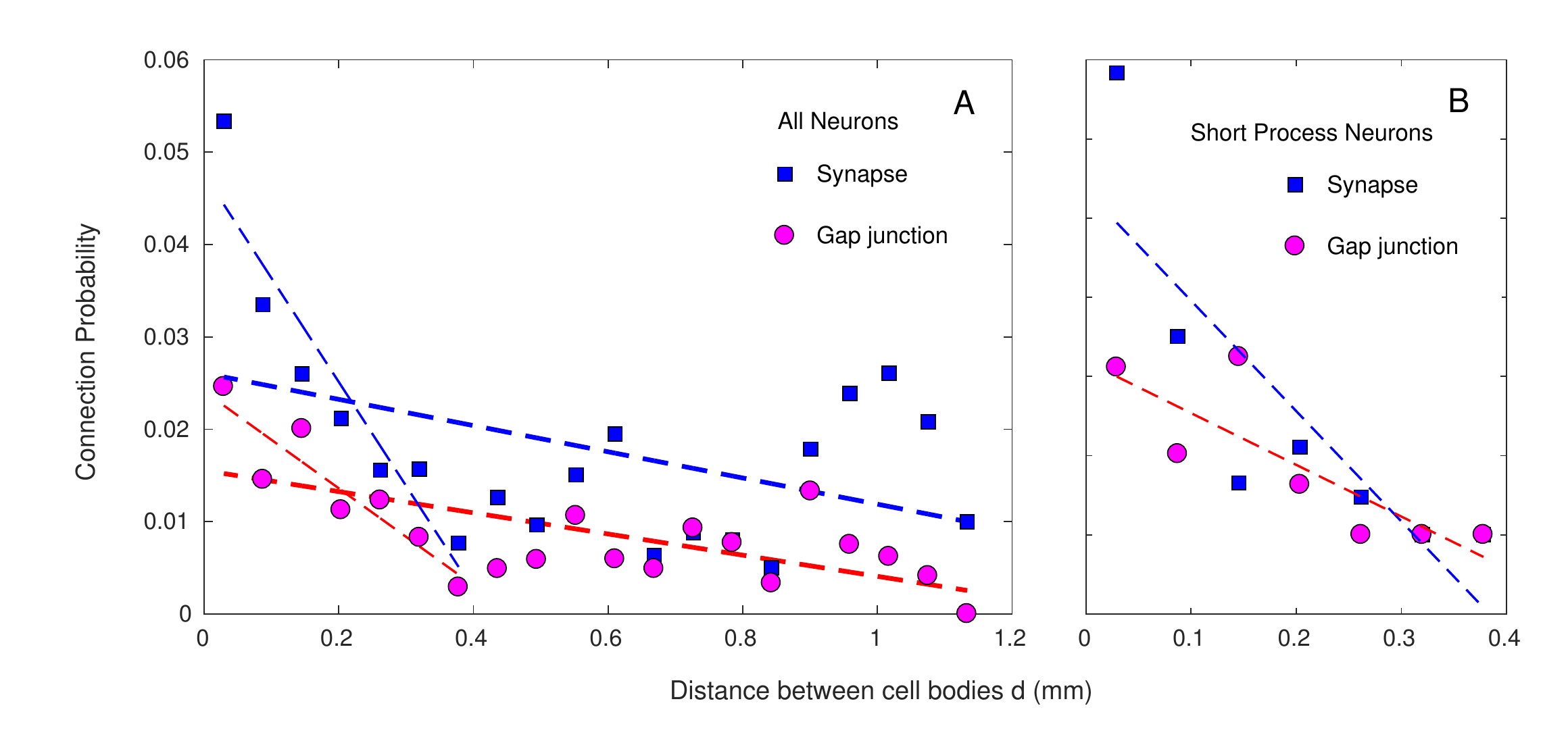}}
\caption{{\bf Dependence of the probability of connection between two
neurons on the physical distance between their cell bodies.}
(A) The variation of the probability of connection between two neurons 
by either synapse (squares) or gap-junction (circles) as a function
of the physical distance between their cell bodies $d$ (measured in mm).
Linear fitting of the functions show a decay with $d$ overall (thick 
broken lines), but the relation is much weaker compared to that
seen between probability of synaptic connection between
two cells and their lineage distance $l$ [see Fig.~\ref{fig2} (D) in
main text]. In particular, the correlation is diluted by the 
relatively high probability for 
synapses to form between neurons whose cell bodies are located at the opposite
ends of the worm (corresponding to the peak around $x=1$ mm).
However, when we focus only on connections between cell bodies
that are in close physical proximity ($d < 0.4$ mm), the dependence on $d$
appears to be much more prominent (thin broken lines). This
stronger correlation between connection probability and $d$ at short 
distances is not necessarily an outcome of constraints imposed by the
process lengths of the neurons. This is suggested by panel (B), where we 
focus exclusively
on neurons with short processes. (B) The relation between 
connection probability between neurons, both of which have 
short processes, and the distance between their cell bodies, $d$, is 
seen to be not more prominent than that already 
seen for all neurons [in panel (A)]. 
}
\label{S3}
\end{figure}

\begin{figure}[h!]
\centerline{\includegraphics[width=0.6\columnwidth]{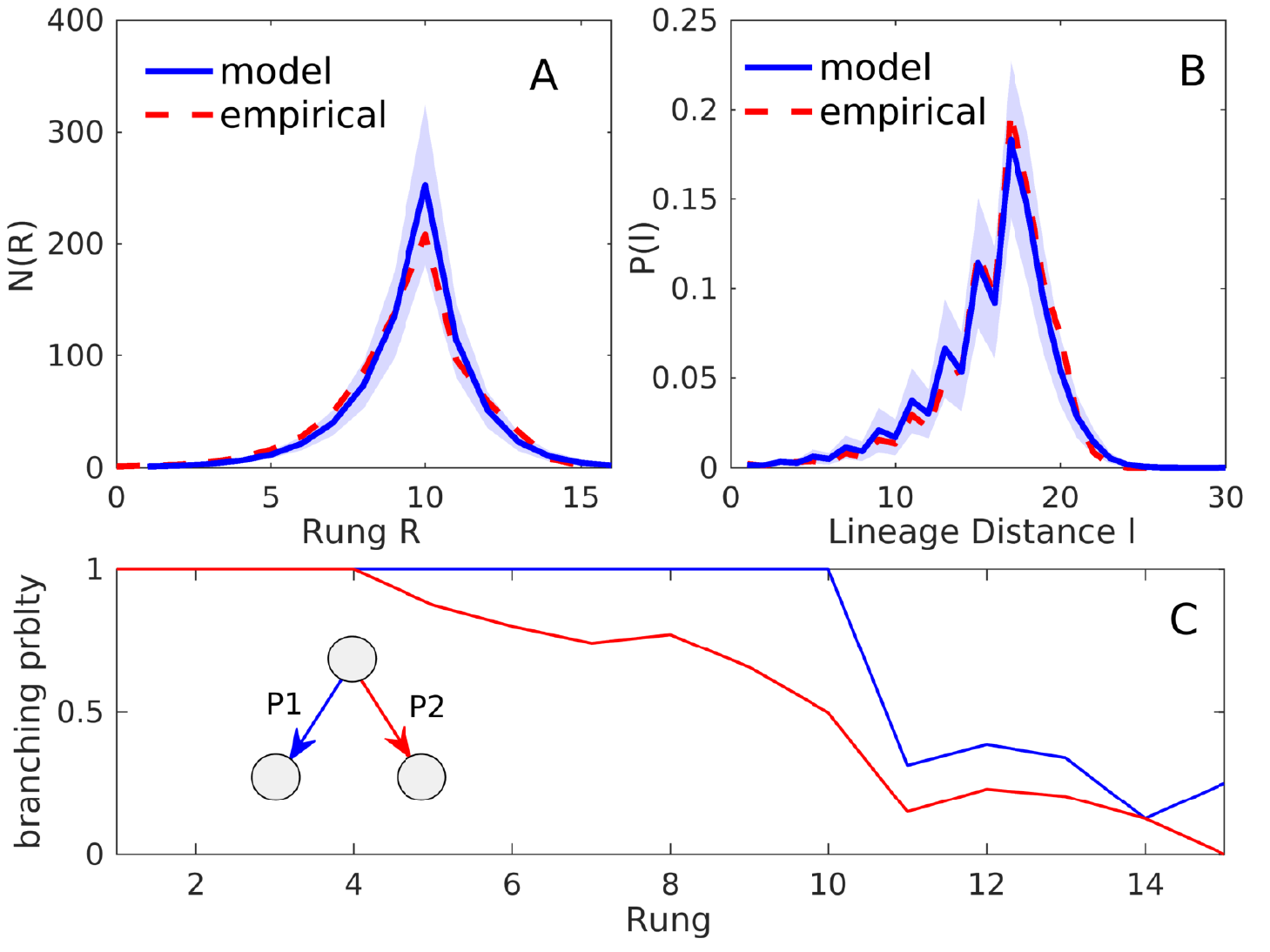}}
\caption{\textbf{A stochastic branching model for the lineage 
tree of cells involved in the development of the {\em C. elegans} 
somatic nervous system.}
(A) Comparison of the distribution of rung $R$ occupied by each cell (progenitor
cells of the neurons, as well as, differentiated neurons) in the lineage
tree obtained empirically (broken curve) with that generated by the model 
(solid curve shows the mean computed over an ensemble of $10^3$ realizations,
the dispersion being indicated by the shaded area). 
(B) Comparison of the distribution of lineage distance $l$ between pairs
of differentiated neurons of {\em C. elegans} (broken curve) with
that obtained from the model (solid curve showing the mean computed 
over an ensemble of $10^3$ realizations,
the dispersion being indicated by the shaded area).
The high degree of overlap between the empirical and simulated distributions
indicates that the stochastic branching model is a reasonably accurate 
description of the lineage tree of neurons. 
(C) The branching probabilities
$P1$ (blue curve) and $P2$ (red curve) of a progenitor cell at each rung, estimated
from the empirical lineage tree (by definition, $P1 \geq P2$). 
Note that both of the branching probabilities show a prominent dip 
after rung 10. Guided by this, in the stochastic branching model,
$P1,P2$ have been chosen to have a constant high value upto rung 10
(viz., $P1 = 1$, $P2=0.85$), after which both are decreased to a constant
low value (viz., $P1=0.25$, $P2=0.2$). The inset shows a schematic
of the stochastic branching model where a node, occurring at any rung,
can branch (or not) based on the probabilities $P1$ and $P2$ which will
result in any one of the following three possibilities: 
(i) proliferation occurs along both branches,
(ii) only one branch appears (the other branch leading to either apoptosis
or a non-neural cell fate), and,
(iii) there is no branching so that we obtain a terminal node 
of the tree (i.e., the cell differentiates into a
neuron).
}
\label{S4}
\end{figure}

\begin{figure}[htb]
\centerline{\includegraphics[width=\columnwidth]{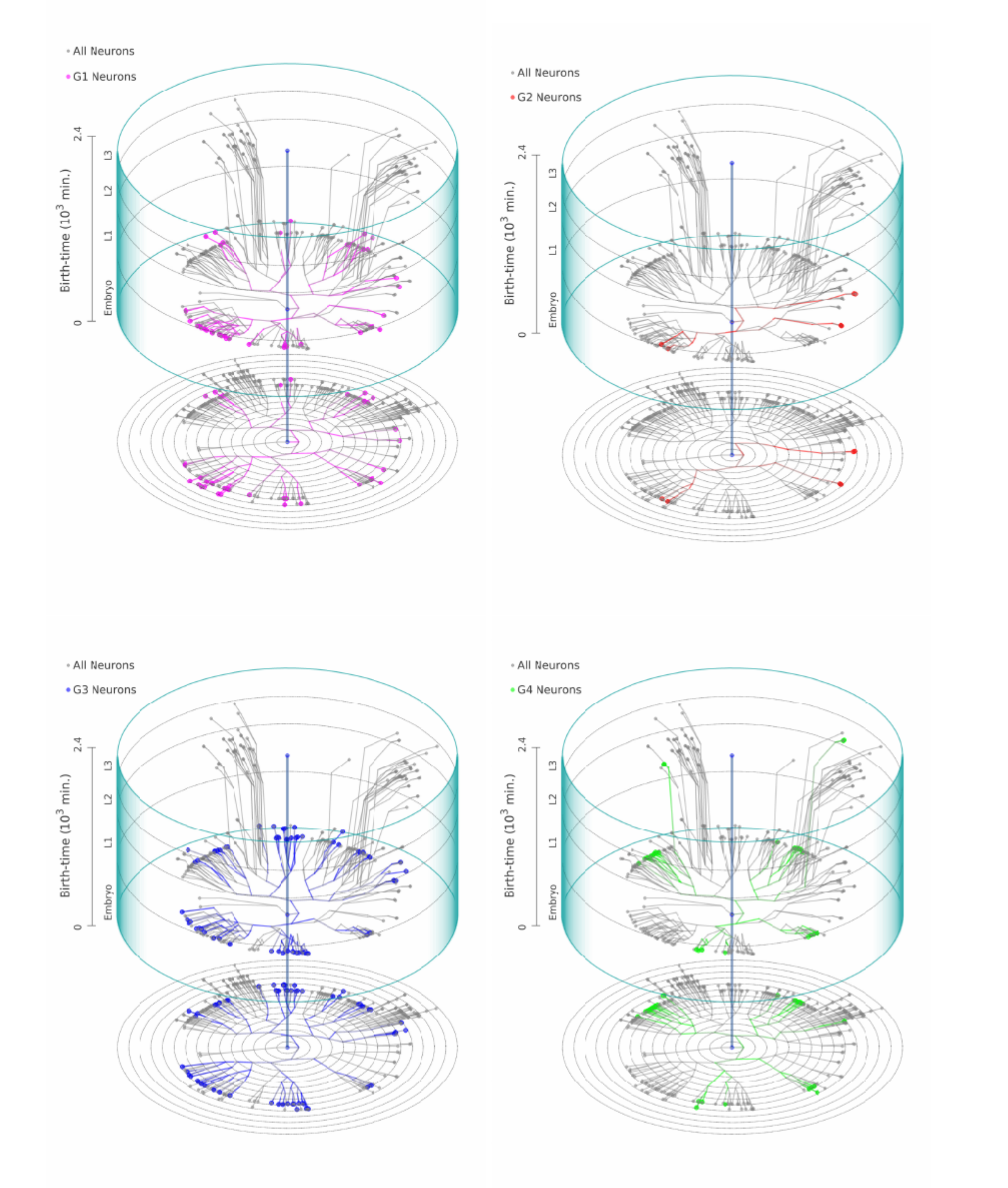}}
\caption{\textbf{Developmental chrono-dendrograms for the Anterior (G1, top left), 
Dorsal (G2, top right), Lateral (G3, bottom left) and Ventral (G4, bottom right) ganglia, showing that each
comprises multiple localized clusters of neurons.}
Colored nodes represent neurons belonging to the specified ganglion
while grey nodes show other neurons. 
Branching lines trace all cell divisions starting from the 
single cell zygote (located at the origin) and terminating at each differentiated neuron. 
The time and rung of 
each cell division is indicated 
by its position along the vertical and radial axis respectively.    
The entire time period is 
divided into four stages, viz., Embryo (indicated as E), L1, L2 and L3. 
A planar projection at the base of each cylinder shows 
the rung (concentric circles)
of each progenitor cell and differentiated neuron.
}
\label{S5}
\end{figure}

\begin{figure}[htb]
\centerline{\includegraphics[width=0.9\columnwidth]{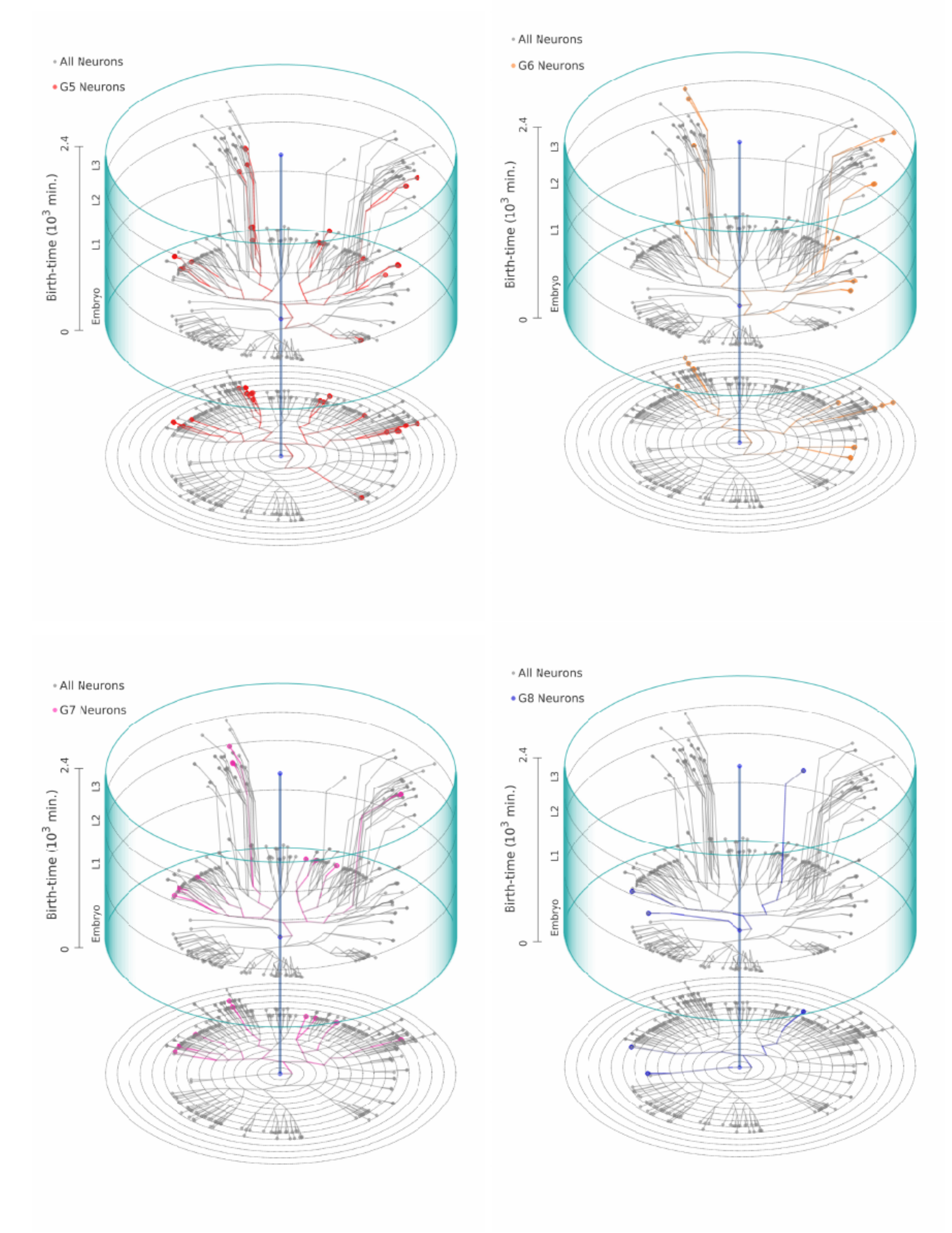}}
\caption{\textbf{Developmental chrono-dendrograms for the Retrovesicular
(G5, top left), Posterolateral (G6, top right), Preanal (G7, bottom left)
and Dorsorectal (G8, bottom right) ganglia, showing that each
comprises multiple localized clusters of neurons.} 
Colored nodes represent neurons belonging to the specified ganglion
while grey nodes show other neurons. 
Branching lines trace all cell divisions starting from the 
single cell zygote (located at the origin) and terminating at each differentiated neuron. 
The time and rung of 
each cell division is indicated 
by its position along the vertical and radial axis respectively.    
The entire time period is 
divided into four stages, viz., Embryo (indicated as E), L1, L2 and L3. 
A planar projection at the base of each cylinder shows 
the rung (concentric circles)
of each progenitor cell and differentiated neuron.
}
\label{S6}
\end{figure}

\begin{figure}[htb]
\centerline{\includegraphics[width=\columnwidth]{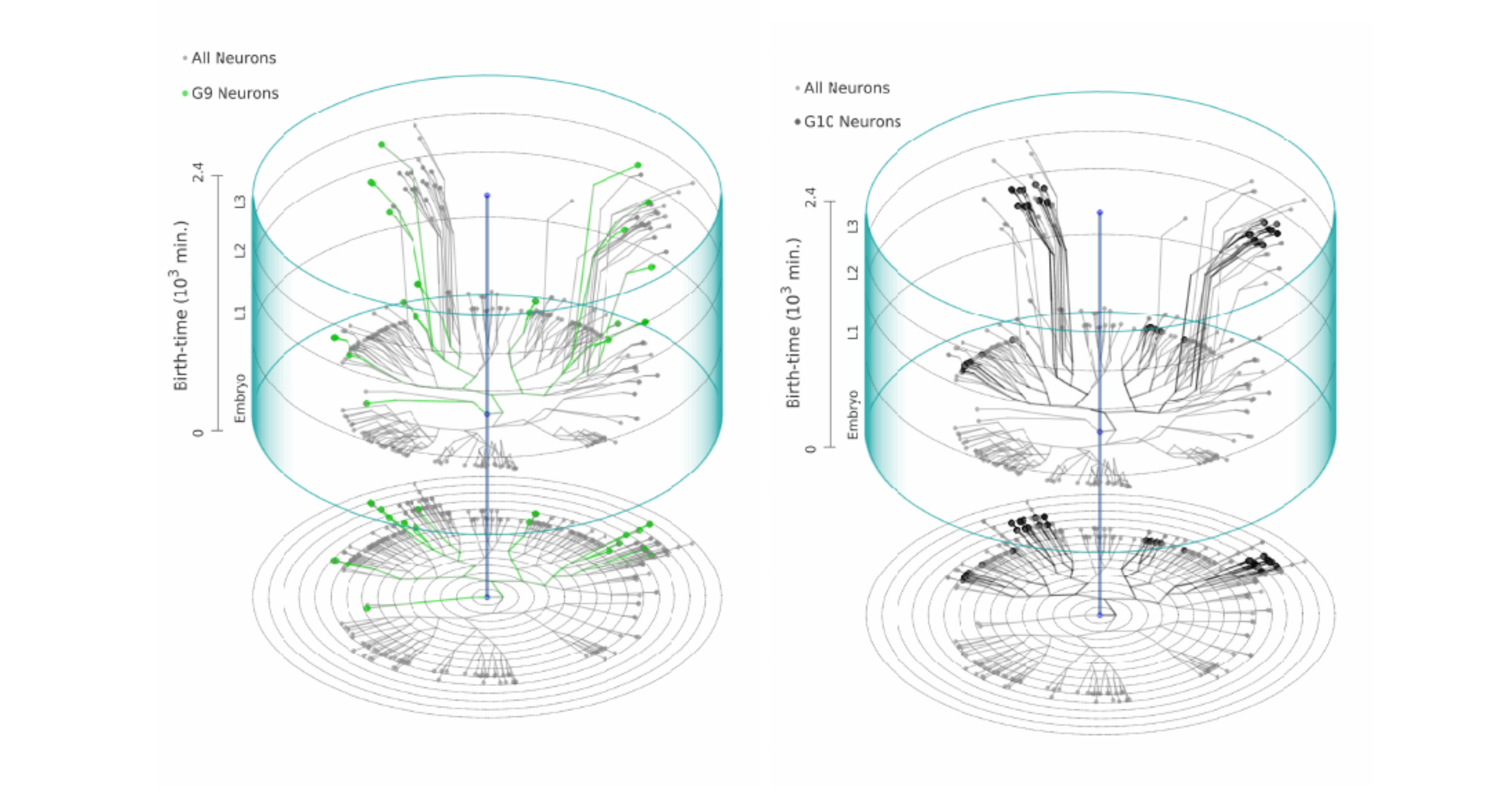}}
\caption{\textbf{Developmental chrono-dendrograms for the Lumbar ganglion
(G9, left) and the Ventral Cord (G10, right), showing that each comprises multiple localized clusters of neurons.} 
Colored nodes represent neurons belonging to the specified ganglion
while grey nodes show other neurons. 
Branching lines trace all cell divisions starting from the 
single cell zygote (located at the origin) and terminating at each differentiated neuron. 
The time and rung of 
each cell division is indicated 
by its position along the vertical and radial axis respectively.    
The entire time period is 
divided into four stages, viz., Embryo (indicated as E), L1, L2 and L3. 
A planar projection at the base of each cylinder shows 
the rung (concentric circles)
of each progenitor cell and differentiated neuron.
}
\label{S7}
\end{figure}

\begin{figure}[h!]
\centerline{\includegraphics[width=0.95\columnwidth]{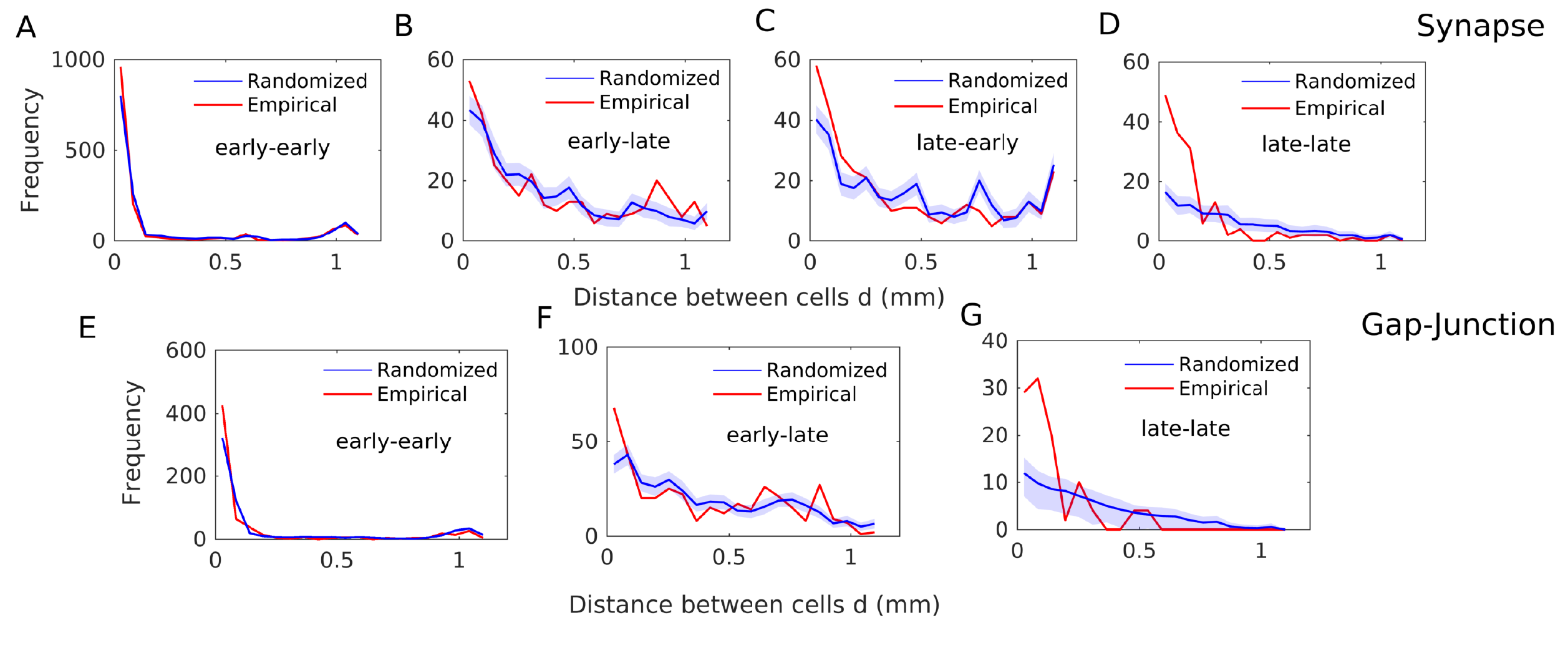}}
\caption{\textbf{
Connections between neurons born at different developmental epochs
are over-represented when the cell bodies are far apart, suggesting
the presence of active processes facilitating such links.
}
Frequency distributions of the distance $d$ between cell bodies of
all neuronal pairs that are connected via synapses (A-D) or 
gap-junctions (E-G).
The distributions for the empirical network (shown in red) are
compared with distributions obtained from surrogate ensembles of
randomized networks (blue curve shows the average over $100$ realizations,
the dispersion being indicated by the shaded area).
The latter are constructed from the empirical network by randomly
rewiring the connections while keeping the total number of connections (degree)
for each neuron, the spatial location of its cell body and its process
length unchanged. In addition, to allow only physically possible connections
between neurons, we have imposed process-length constraint which disallow
linking two cells if the distance between their cell bodies is greater than
the sum of their individual process lengths.
The different panels correspond to the situations where
(A,E) both cells in a connected pair are born in the early developmental
burst, (B,C,F) one is born early and the other is born late [in (B)
it is the pre-synaptic neuron which is born early, while in (C) the
post-synaptic neurons appears in the early developmental burst], and 
(D,G) both cells are born late.
When two neurons are born in the same developmental epoch (either early or
late), the empirical frequency distribution is seen to have significantly 
higher values than the randomized distribution at low $d$ (seen in
panels A and D, and even more prominently in panels E and G), 
indicating that neurons prefer to connect
to other members of their birth cohort whose cell bodies are in close
proximity. This is particularly evident for neurons born in the
late developmental epoch. Note that this result complements
the earlier observation that birth cohort homophily is seen specifically
for neurons whose cell bodies are located relatively close to each other
(Fig.~\ref{figS1}). 
More intriguingly, connections between neurons whose cell bodies lie 
far apart are seen to occur more frequently than expected by chance
when the pre-synaptic neuron is born early and the post-synaptic neuron
is born late (see panel B). A similar phenomenon is also seen in the case
of early- and late-born neurons connected by gap junctions (see panel F).
These results suggest the presence of an active process forming 
connections between neurons born in different epochs.}
\label{S8}
\end{figure}

\begin{figure}[h!]
\centerline{\includegraphics[width=0.95\columnwidth]{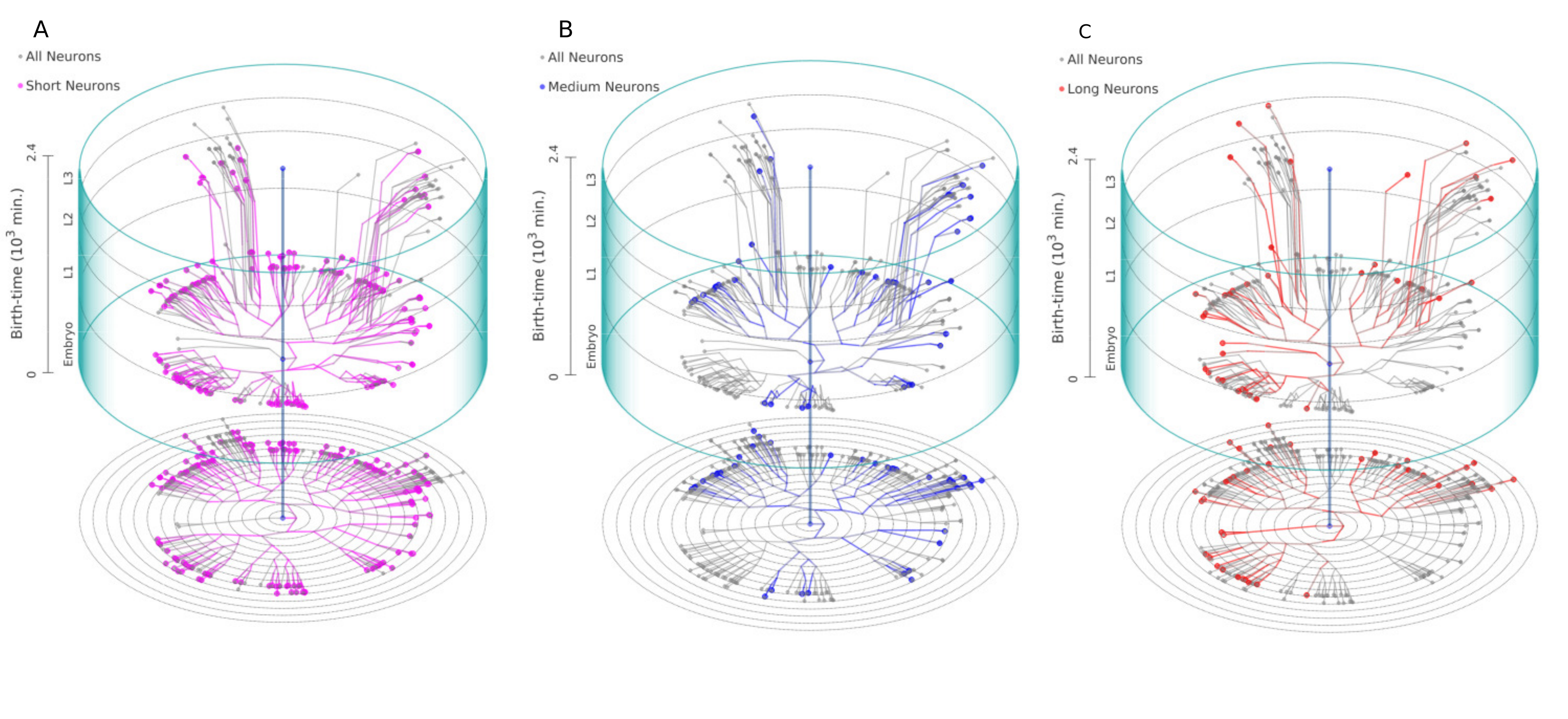}}
\caption{\textbf{Absence of segregated clusters in the developmental 
chrono-dendrograms for neurons having similar process lengths (viz., short,
medium and long) suggest that process length is not exclusively
determined by lineage.}
Colored nodes represent neurons having a specified process length,
viz., short in (A), medium in (B) and long in (C),
while grey nodes show other neurons. 
Branching lines trace all cell divisions starting from the 
single cell zygote (located at the origin) and terminating at 
each differentiated neuron. 
The time and rung of 
each cell division is indicated 
by its position along the vertical and radial axis respectively.    
The entire time period is 
divided into four stages, viz., Embryo (indicated as E), L1, L2 and L3. 
A planar projection at the base of each cylinder shows 
the rung (concentric circles)
of each progenitor cell and differentiated neuron.
}
\label{S9}
\end{figure}

\begin{figure}[h!]
\centerline{\includegraphics[width=0.95\columnwidth]{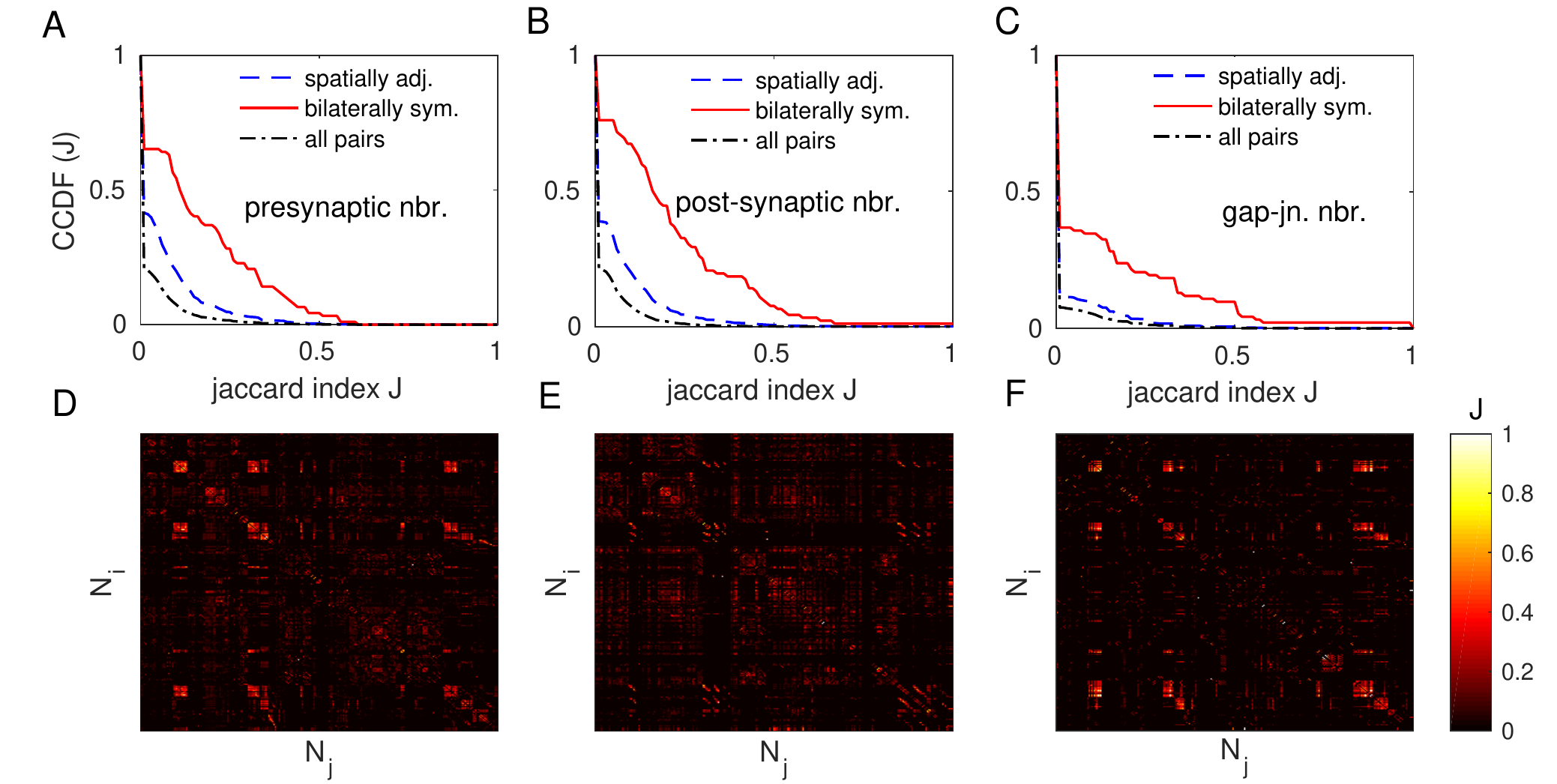}}
\caption{\textbf{
Physical proximity alone cannot explain the high degree of overlap
between the cells that each member of a bilaterally symmetric pair of
neurons connect to.}
Complementary cumulative probability distributions (CCDF) of the overlap
between the sets
$\mathcal{N}_G (i)$, $\mathcal{N}_G (j)$ of the
neighbors (defined for a network $G$) of neurons $N_i$ and $N_j$, where
the indices $i$ and $j$ can run over (i) all pairs of neurons (dash-dotted
curves), (ii) only bilaterally symmetric pairs (solid curves) or 
(iii) pairs whose
cell bodies are spatially adjacent to each other ($d<0.05$ mm, broken curves).
The overlap is measured in terms of the Jaccard index $J$, defined for
the pair $i,j$ as 
$J (i,j) = [\mathcal{N}_G (i) \cap \mathcal{N}_G (j)]/[\mathcal{N}_G (i) \cup \mathcal{N}_G (j)]$,
where $\cap$ and $\cup$ refers to intersection and union of two sets,
respectively.
The different panels correspond to different networks $G$ used to define
neighbors for a neuron, viz., (A) pre-synaptic neighbors, i.e., cells
from which the neuron receives a synaptic connection, (B) post-synaptic 
neighbors, i.e., cells to which the neuron sends a synaptic connection,
and (C) gap-junctional neighbors, i.e., cells to which a neuron is coupled
via a gap junction.
We note that the overlaps between the neighborhoods (for all three types of
network neighbors considered here) of 
bilaterally symmetric neurons are consistently higher than that
of all pairs of neurons, as well as, of pairs whose cell bodies are in 
close physical proximity.
Thus, bilaterally symmetric neurons share neighbors to a much greater extent
than that expected by their cell bodies being located
close to each other.
(D-F) The Jaccard index matrices $J$ showing overlaps between the 
neighbors for every pair of neurons $N_i$, $N_j$ when the network
neighborhood defined is that of (D) pre-synaptic partners,
(E) post-synaptic partners and (F) gap-junctional partners. The large overlap 
between neighbors of bilaterally
symmetric neurons is indicated by the occurrence of bands of brightly 
colored entries along the diagonal (note that bilaterally symmetric neurons are
always located on adjacent rows/columns). 
}
\label{S10}
\end{figure}

\begin{figure}[h!]
\centerline{\includegraphics[width=\columnwidth]{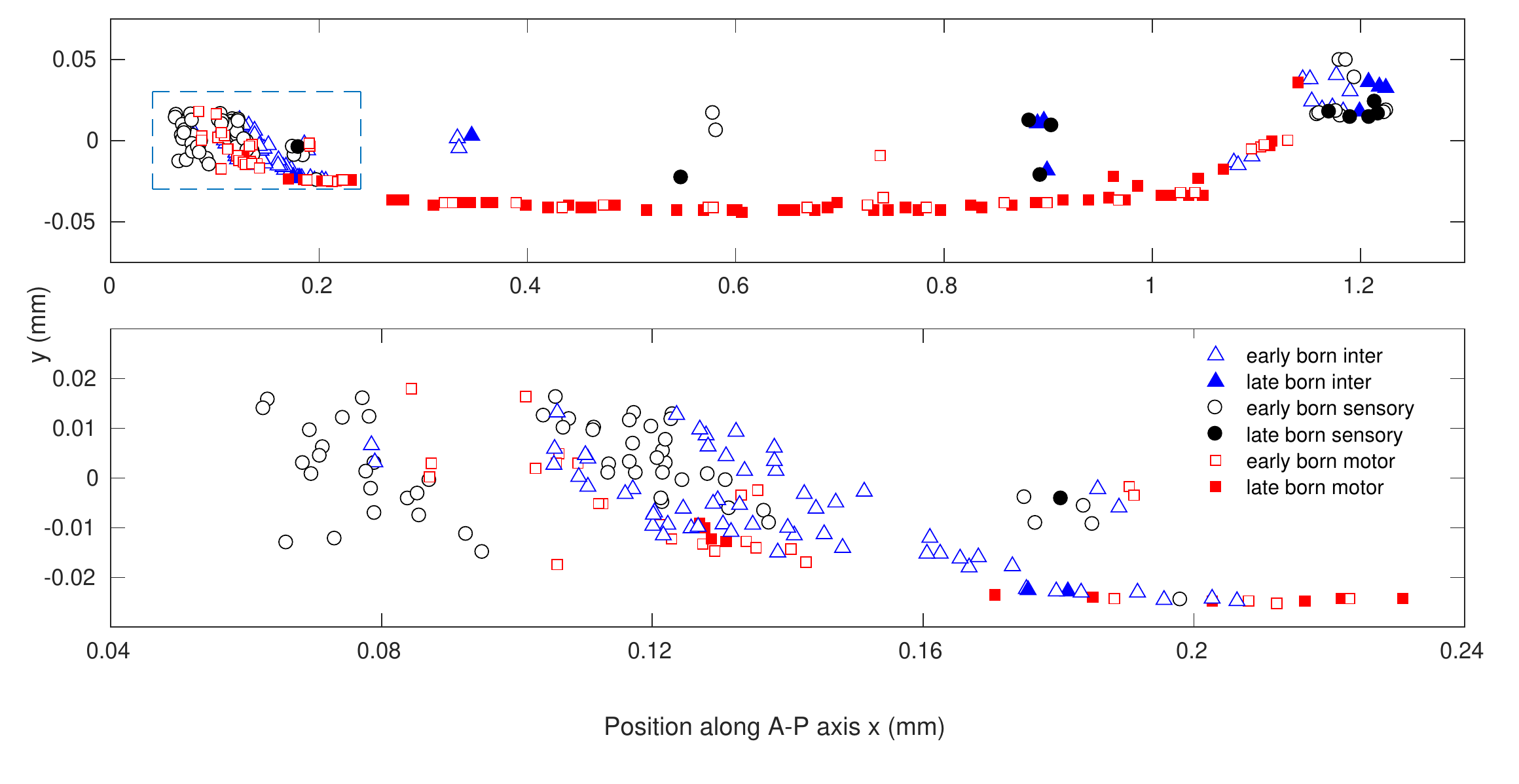}}
\caption{\textbf{Spatial distribution of the cell bodies of sensory,
inter and motor neurons of the somatic nervous system of 
{\em Caenorhabditis elegans.}} 
Projections of the physical locations of the neuronal cell bodies, 
distinguished according to functional type (sensory: circles, inter:
triangles and motor: squares) 
and whether they appear
in the early (unfilled symbols) or late (filled symbols) 
developmental epochs, on the two-dimensional plane 
formed by the anterior-posterior (AP) axis ($x$, along the horizontal)
and the dorsal-ventral axis ($y$, along the vertical). 
Top panel shows the entire worm, with its body oriented such that the head
is located near the
left end and tail near the right end of the plane. The bottom panel shows
a magnified view of the region near the head (bounded by broken lines
in the top panel). We note that almost all cells in this region appear
at the embryonic stage, during the early burst of development.
By contrast, the ventral cord predominantly comprises neurons that 
appear in the post-embryonic stage (see top panel).
}
\label{S11}
\end{figure}

\begin{figure}[h!]
\centerline{\includegraphics[width=\columnwidth]{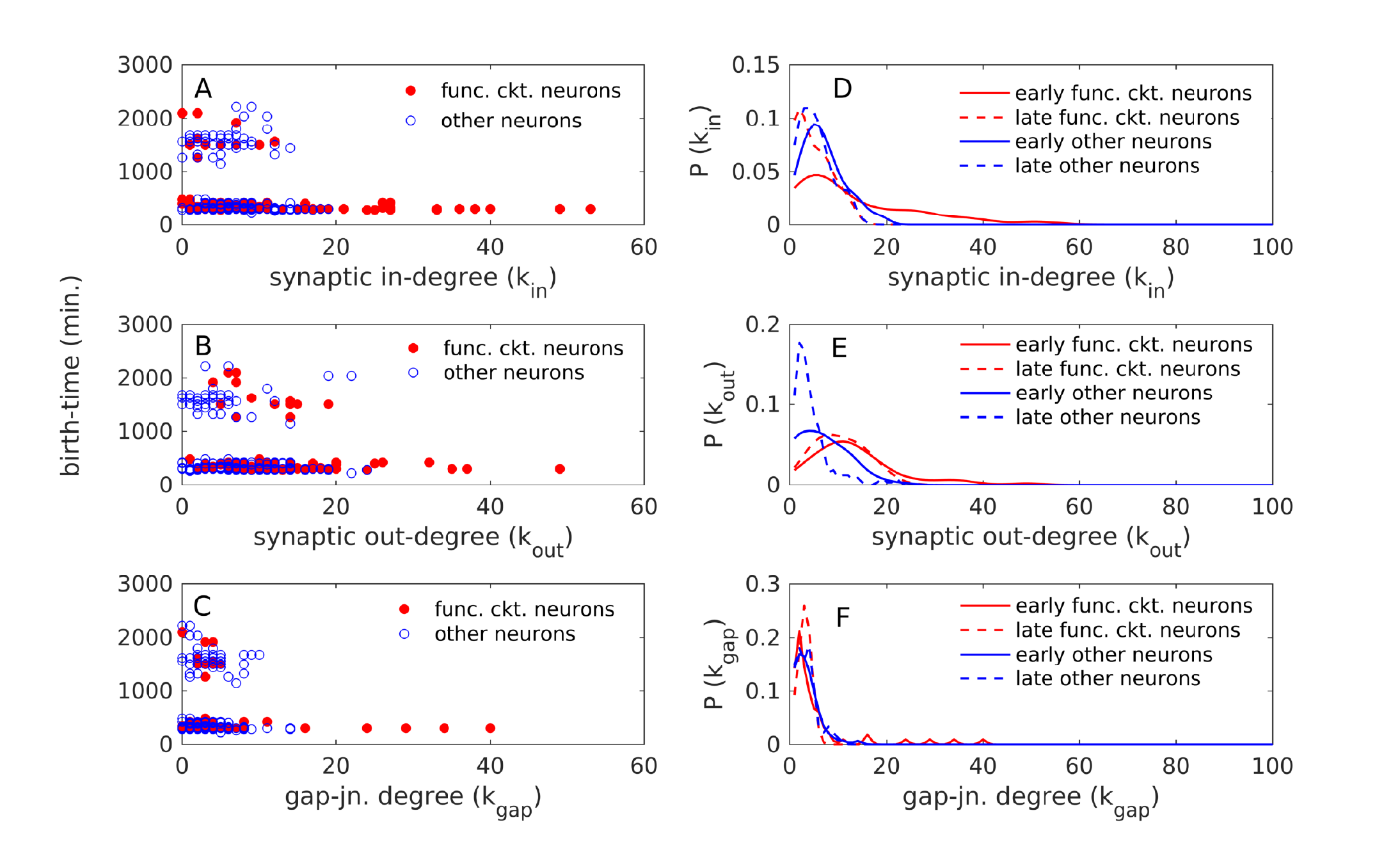}}
\caption{\textbf{The number of synaptic connections of a neuron is influenced 
by its functional criticality, as well as, the developmental epoch in which 
it appeared.}
(A-C) Scatter plots indicating the relation between the time of
appearance of a neuron and its number of (A) incoming synaptic connections 
from other cells (synaptic in-degree), (B) outgoing synaptic connections 
to other cells (synaptic out-degree) and (C) gap junctions with other cells
(gap-junctional degree).
Filled circles represent neurons belonging 
to any of seven previously identified functional circuits (see
Fig.~\ref{fig8} in main text) while unfilled circles show other neurons. 
(D-F) Probability distributions of different types of connections
for neurons categorized in terms of those which are functionally critical,
i.e., belong to a functional circuit (red), or not (blue), 
and whether they appear in the
early (solid curve) or late (broken curve) developmental epochs.
The different panels correspond to (D) synaptic in-degree, 
(E) synaptic out-degree and (F) gap-junctional degree. 
Synaptic in-degree for functionally critical, 
early-born neurons is seen to have a heavy-tailed distribution which
is significantly different from that of the
other types of neurons (in terms of a $2$-sample Kolmogorov-Smirnov 
test at $5\%$ level of significance).
The heavy tail arises from the appearance of
a set of neurons having exceptionally high in-degree [the filled
circles having $k_{in}>20$ shown in panel (A)], that play the role
of connector hubs in the nervous system. These correspond
to the early born R6 category neurons [see Fig.~\ref{fig7}~(B)]. 
On excluding these, the synaptic in-degree distributions
of the early-born functional circuit neurons become indistinguishable from that
of other categories of neurons.
For the case of synaptic out-degree, however, the distributions for functionally
critical neurons that are born at different epochs are statistically 
indistinguishable. However, for other neurons, the distribution of those
that are born in the later, post-embryonic developmental burst are distinct
from those that are born early
(demonstrated by a $2$-sample 
Kolmogorov-Smirnov test at $5\%$ level of significance). 
This statistically significant difference between the outgoing 
connections of early and late-born neurons could arise from the former
neurons being present for a much longer period during which they can 
send out synapses.
Distributions of gap junctional connections for all categories of 
neuron appear to be statistically indistinguishable, suggesting that
gap junction formation is relatively unaffected by the functional
criticality or time of appearance of the neurons.
}
\label{S12}
\end{figure}

\end{document}